\documentclass[smallextended, a4paper]{svjour3}

\usepackage{amsmath}
\usepackage{ifxetex}

\makeatletter
\let\cl@chapter\undefined
\makeatletter

\ifxetex
\usepackage{fontspec}
\usepackage{unicode-math}
\newcommand\psiCompat{{\normalfont ψ}}
\else
\usepackage{mathptmx}
\usepackage{amssymb}
\usepackage{wasysym}
\usepackage{stmaryrd}
\usepackage[utf8]{inputenc}
\usepackage[T1]{fontenc}
\newcommand\coloneq{:=}
\newcommand\Coloneq{::=}
\newcommand\mdlgwhtcircle{\ensuremath{\bigcirc\,}}
\newcommand\mdlgwhtdiamond{\lozenge\,}
\newcommand\mdlgwhtsquare{\square\,}
\newcommand\psiCompat{$\varphi$}
\DeclareUnicodeCharacter{03C8}{\ensuremath{\varphi}}
\fi

\usepackage{multirow,booktabs}

\usepackage{tikz}
\usetikzlibrary{positioning,shapes.geometric,calc}
\usepackage[noliterate]{maude}
\usepackage[colorlinks, citecolor=teal]{hyperref}
\usepackage[capitalise,noabbrev]{cleveref}
\usepackage{bussproofs}
\usepackage{breakcites}

\newcommand*\N{\mathbb{N}}
\newcommand\ctlNext{\ensuremath{\mdlgwhtcircle\,}}
\newcommand\ctlAllw{\ensuremath{\mdlgwhtsquare\,}}
\newcommand\ctlEvly{\ensuremath{\mdlgwhtdiamond\,}}
\newcommand\ctlUntil{\ensuremath{\,\mathbf{U}\,}}
\newcommand*\kywd[1]{\texttt{\bfseries #1}}
\newcommand*\skywd[1]{{\normalfont\texttt{\color{darkgray}\bfseries #1}}}
\lstnewenvironment{maudexec}[1][]{\lstset{language={}, #1}}{}
\newcommand\idle{{\normalfont\skywd{idle}}}
\newcommand\fail{{\normalfont\skywd{fail}}}
\newcommand\seq{{\normalfont\texttt;}}
\newcommand\disj{{\normalfont\texttt|}}
\newcommand\ifthel[3]{{\normalfont \,#1\, \texttt? \,#2\, \texttt: \,#3}}
\newcommand\cond[2]{{\normalfont\skywd{match}\; #1 \;\skywd{s.t.}\; #2}}
\newcommand\opsem{\twoheadrightarrow}
\newcommand\xs{\ensuremath{\mathcal{X\!S}}}
\newcommand\cterm{\mathrm{cterm}}
\newcommand*\ao{\,\lower1pt\hbox{@}\,}
\newcommand\subterm{\mathrm{subterm}}
\newcommand\rewcond{\mathrm{rewc}}
\newcommand\vctx{\mathrm{vctx}}

\newcommand*\ttrew{\;\text{\tt=>}\;}
\newcommand*\matchrewf{\skywd{matchrew} \;P\; \skywd{s.t} \;C \; \skywd{by} \; x_1 \;\skywd{using} \; \alpha_1 , \; \ldots , x_n \; \skywd{using} \; \alpha_n}

\newcommand*\ltssl{\ensuremath{\mathcal O}}
\newcommand*\ltmsl{\ensuremath{\mathcal M_{\alpha, t}}}

\journalname{Automated Software Engineering}

\title{Model checking strategy-controlled systems \\ in rewriting logic}
\author{Rubén Rubio \and Narciso Martí-Oliet \and Isabel Pita \and Alberto Verdejo}
\institute{Facultad de Informática, Universidad Complutense de Madrid, Spain,\\
\email{rubenrub@ucm.es (R. Rubio, corresponding author), narciso@ucm.es (N. Martí-Oliet), ipandreu@ucm.es (I. Pita), jalberto@ucm.es (A. Verdejo) }.\\
ORCID: 0000-0003-2983-3404 (R. Rubio), 0000-0002-6576-762X (N. Martí-Oliet), 0000-0003-4915-5452 (I. Pita), 0000-0002-7374-3214 (A. Verdejo).}
\date{\today}

\begin{document}

\maketitle

\begin{abstract}
	Rewriting logic and its implementation Maude are an expressive framework for the formal specification and verification of software and other kinds of systems. Concurrency is naturally represented by nondeterministic local transformations produced by the application of rewriting rules over algebraic terms in an equational theory. Some aspects of the global behavior of the systems or additional constraints sometimes require restricting this nondeterminism. Rewriting strategies are used as a higher-level and modular resource to cleanly capture these requirements, which can be easily expressed in Maude with an integrated strategy language. However, strategy-aware specifications cannot be verified with the builtin LTL model checker, making strategies less useful and attractive.
In this paper, we discuss model checking for strategy-controlled systems, and present a strategy-aware extension of the Maude LTL model checker. The expressivity of the strategy language is discussed in relation to model checking, the model checker is illustrated with multiple application examples, and its performance is compared.
\end{abstract}

\section{Introduction}

	Rewriting logic~\cite{rewritingLogic,20years} is a natural and expressive framework for the formal specification and analysis of concurrent systems and logics. Their objects are described using arbitrary signatures where terms are considered modulo equations and structural axioms, their state transitions are expressed using rewriting rules, and their executions are the successive and independent application of these rules. In each step of the process, the rule, the position where it is applied, and the matching substitution are nondeterministically chosen, yielding potentially many evolutions of the system. The spatial and temporal locality of rules is the cornerstone of the natural and simple representation of concurrency and deduction, but it is sometimes convenient to tame this nondeterminism to capture the global behavior of the system or other specific restrictions.
This is the purpose of strategies, which have aroused interest since the introduction of the $\lambda$-calculus~\cite{barendregt} and have been profusely studied in the context of rewriting and reduction~\cite{allthat,terese,extstrat}, as well as in artificial intelligence~\cite{heuristics}, automated deduction~\cite{satstrats}, game theory~\cite{gameTheoryStrat}, computational chemistry~\cite{chemicalStrat}, etc.
Regarding modeling and formal specification, strategies are useful to separate the basic rules of the model behavior from its control, following the well-know sofware engineering principles of separation of concerns~\cite{separationConcerns}, modularity, abstraction, and incremental development. This idea is enunciated in the Kowalski's motto \emph{Algorithm = Logic + Control}~\cite{kowalski} and developed in the Lescanne's \emph{Rule + Control} approach~\cite{lescanneOrme}, arguing that ``computer programs would be more often correct and more easily improved and modified if their logic and control aspects were identified and separated in the program text''. For example, the terms and deduction rules of an inference system can be expressed as a rewrite theory and be proven sound, but only a careful application of these rules will efficiently lead to the desired deductions.
This approach has given place to various executable strategy languages like ELAN~\cite{elan}, TOM~\cite{tom}, Stratego~\cite{stratego} for program transformation, $\rho$Log~\cite{rholog}, \textsc{Uppaal Stratego}~\cite{uppaalStratego}, and more recently Porgy~\cite{porgyJournal} for graph rewriting. Unlike the strategies usually considered for the $\lambda$-calculus and abstract rewriting, these strategies are syntactically represented as programs and the next steps are not only dependent on the last state but may depend on the whole history of the derivation. These languages have been applied to several real problems, among others~\cite{porgy3NF,porgyFinantial,srewSocialNetworks,chemicalStrat,rhologAbac}.

	Maude~\cite{maude,allmaude} is a specification language based on rewriting logic and an interpreter that allows executing and analyzing its specifications. Maude includes a strategy language for controlling the rewriting process~\cite{towardsStrategy}, maintaining a separation between rules and strategies, so that different strategies can be compositionally specified to easily control the same rewriting system.
The Maude strategy language has been used to specify semantics of programming languages like Eden~\cite{eden}, biologically-inspired computational models~\cite{memstratmc}, neural networks~\cite{neuralNetworks}, and many more~\cite{sudoku,completion,pssm,ambientCalculus}.
However, while it is easy to check properties on pure rule-based specifications using the LTL model checker included in Maude~\cite{maudemc}, this was no longer possible for systems specified with strategies, as pointed out by some authors~\cite{membrane}. In order to solve this problem, we have extended here the builtin Maude LTL model checker for systems controlled by strategies.

	In this paper, we discuss model checking for strategy-controlled systems against any linear-time logic that is well defined in the uncontrolled system. The main intuition is that properties should only be checked in the subset or subtree of executions allowed by the strategy.
This idea is already present in the \emph{strategic logics}~\cite{mogaveroJournal}, where strategies are part of the property specification instead of the system, and in \textsc{Upaal Stratego}~\cite{uppaalStratego} for simpler memoryless strategies, where the selected subset of executions is called \emph{strategy space}.
We show that a general procedure for actual model checking using the standard algorithm for the desired logic is transforming the model so that it incorporates the restrictions imposed by the strategy. In order to apply this approach to the Maude strategy language, we provide it with a small-step operational semantics to precisely determine which are the executions described by a strategy expression and construct the transformed model. The expressivity of the language and conditions for model checking to be decidable are discussed too.
In coherence with these ideas and with the semantics, we have implemented an extension of the Maude LTL model checker to deal with strategy-controlled systems, which has already been given various applications~\cite{bitmlmc,memstratmc,metatrans}.

	This article extends the conference paper \cite{fscd} including some advancements in \cite{btimemc} with an improved and systematic presentation, new results, further details and proofs, performance comparisons, and more examples. \Cref{sec:preliminars} reviews some precedents and well-known topics that are required to follow the rest of the paper. \Cref{sec:smc} defines and discusses the model-checking problem for abstract systems controlled by strategies, which is particularized in~\cref{sec:mcslang} for the Maude strategy language described in~\cref{sec:slang}. \Cref{sec:mcslang} also includes novel results about the expressivity of the strategy language in relation with model checking, and conditions for the decidability of this problem. \Cref{sec:maudesmc} introduces the extension of the Maude LTL model checker, whose implementation is described in~\cref{sec:implementation} and which is evaluated in~\cref{sec:evaluation}. Original examples are shown in~\cref{sec:examples}. All the material, including the model checker, its documentation and source code, the examples described in this paper and many more, is available online~\cite{stratweb}.

\section{Preliminaries} \label{sec:preliminars}

	Let us recall some basic concepts and notation about languages, rewriting logic and model checking, which will be extensively used along the paper. The dining philosophers example introduced in~\cref{sec:maude} will be the running example in the rest of the paper.

\subsection{Languages and automata over finite and infinite words} \label{sec:languages}

	Some basic knowledge about the theory of formal language is assumed, but we review the notation for the set $\Sigma^*$ of all finite words on the alphabet $\Sigma$, whose subsets are called languages, for the length of a word $|w|$, and for the operations on languages like union $L \cup M$, intersection $L \cap M$, concatenation $LM$, power $L^n$, and the Kleene star $L^* = \cup_{n \in \N} L^n$. We also write $w_k$ for the $k$-th symbol of a word $w \in \Sigma^*$ starting from zero, $w^k$ for the suffix starting at index $k$, and $w^{< k}$ for the prefix of length $k$.
Infinite words~\cite{omegaLanguages,infiniteWords} are infinite sequences $w : \N \to \Sigma$ of symbols, and languages over infinite words are subsets of the set of all such words $\Sigma^\omega$, whose typical operations are well defined unless  concatenation after an infinite word is involved. Moreover, the infinite concatenation of a finite-word language is written $L^\omega = \{ w_1 w_2 \cdots : w_k \in L \setminus \{\varepsilon\} \}$. Like for classical languages, there is a Chomsky hierarchy of $\omega$-language classes with similar recognizing devices, but $\omega$-regular languages are the most studied because of their application on model checking and the description of reactive systems. They are recognized by Büchi automata, $M = (Q, \Sigma, \delta, q_0, F)$ where $Q$ is a finite set of automaton states, $q_0$ is an initial state, $\delta : Q \times \Sigma \to \mathcal P(Q)$ is a nondeterministic transition function, and $F$ is an \emph{acceptance condition}.\footnote{Büchi automata are very similar to finite automata for regular languages, but, since infinite words do not end, final states are replaced by acceptance conditions. Unlike finite automata, deterministic Büchi automata are less expressive than their nondeterministic counterpart.} A word $w$ is accepted if there is a run $\pi = q_0 q_1 \cdots$ such that $q_k \in \delta(q_{k-1}, w_k)$ satisfying the acceptance condition. A Büchi acceptance condition is a subset $F \subseteq Q$ of states from which at least one must occur infinitely often in the run, i.e.\ $\mathrm{inf}(\pi) \cap F \neq \emptyset$ with $\mathrm{inf}(\pi) = \{ q \in Q : q \text{ appears infinitely often in } \pi \}$.
This class of languages can also be described with $\omega$-regular expressions as in the finite case:
\begin{align*}
	\alpha & \,\Coloneq\, \emptyset \mid \varepsilon \mid s \mid \alpha \alpha \mid (\alpha \mid \alpha) \mid \alpha^* \mid \alpha^\omega
\end{align*}
that are given meaning as $L(\emptyset) = \emptyset$, $L(\varepsilon) = \{ \varepsilon \}$, $L(s) = \{s\}$ for $s \in S$, $L(\alpha\beta) = L(\alpha) L(\beta)$, $L(\alpha \mid \beta) = L(\alpha) \cup L(\beta)$, $L(\alpha^*) = L(\alpha)^*$, and $L(\alpha^\omega) = L(\alpha)^\omega$. These expressions must also obey the restrictions of the $\omega$-language operations.

	Finite and infinite words can be considered together $\Sigma^\infty \coloneq \Sigma^* \cup \Sigma^\omega$, whose languages are named $\infty$-languages.
A prefix $\sqsubseteq$ is naturally defined on $\Sigma^\infty$ where $w \sqsubseteq v$ iff $w = v$ if $w$ is infinite, or otherwise if $v \in \{ w \} \Sigma^\infty$.
An infinite word $w \in \Sigma^\omega$ is an accumulation point of a language $L \subseteq \Sigma^\infty$ if for all $n \in \N$ there is a word in $L$ with $w^{< n}$ as prefix, and $L$ is closed if it contains all its accumulation points. The notion of closed language will appear in some properties of this paper, and as a side note, it coincides with the topological concept of closed set for a topology that is engendered by the chain-complete order $\sqsubseteq$, by a metric $d(w, v) = \min \{ 0, 2^{-n} : w_n \neq v_n \wedge |w| < n \leq |v| \wedge |v| < n \leq |w| \}$, and by other equivalent means.

	Other language classes have also been translated to the infinite word setting including $\omega$-recursively enumerable languages. The analogue of Turing machines are $\omega$-Turing machines with acceptance conditions similar to those of Büchi automata. Language hierarchies and automata are only meaningful for finite alphabets, but the notion of word does not lose sense when the base set is not finite. This circumstance will often happen in the rest of the paper.

\subsection{Strategies} \label{sec:strategies}

	An \emph{abstract reduction systems} (ARS) or \emph{transition system} $\mathcal A = (S, G)$ consists of a set of states $S$ and a binary relation $G \subseteq S \times S$ on them. Sometimes transition systems are labeled $\mathcal A = (S, A, G)$ with an additional set $A$ and $G \subseteq S \times A \times S$. However, we will refer here to unlabeled systems for simplicity, since results can be easily extended to labeled ones.
Arrows are frequently used to write $s \to s'$ instead of $(s, s') \in G$. We say that $s \to s'$ is an \emph{execution step}, that $s'$ is a \emph{successor} of $s$, and that an \emph{execution} in $\mathcal A$ is a finite or infinite sequence of states $s_0 \to s_1 \to \cdots \to s_n$ connected by the relation. They are represented as finite $s_0 s_1 \cdots s_n$ or infinite $s_0 s_1 \cdots$ words, and the sets $\Gamma^*_{\mathcal A} \subseteq S^*$, $\Gamma^\omega_{\mathcal A} \subseteq S^\omega$ and $\Gamma_{\mathcal A} \subseteq S^\infty$ are the finite, infinite and mixed executions, respectively. A subscript $s \in S$ like $\Gamma_{\mathcal A, s}$ indicates that only executions starting at this state are included.

	Transition systems are extensively used for formal modeling in computer science and engineering. Some logics used in the verification of these systems, including those we will describe in~\cref{sec:modelchecking}, only consider nonterminating executions for simplicity~\cite{pneuliLTL}. In many cases, finite executions in $\Gamma^*_{\mathcal A}$ are not meaningful as they do not represent complete executions, but strict prefixes of those. Anyhow, a real system may of course present both finite and infinite executions. The usual convention to solve this problem is the so-called \emph{stuttering extension} that considers valid finite execution as infinite ones by repeating their last state forever. Sometimes this can be implemented in the transition system by adding self-loops to deadlock states, but if the states where the model execution is allowed to halt do not coincide with deadlock states, the following definition is useful.

\begin{definition} \label{def:halting}
Given a transition system $\mathcal A = (S, \to)$ and a set $H \subseteq S$ of halting states, the stuttering extension of $\mathcal A$ with respect to $H$ is
\[ \mathcal A_H = (S \times \{0\} \cup H \times \{1\}, \to_H) \]
where $(s, 0) \to_H (s', 0)$ iff $s \to s'$ for all $s, s' \in S$, and $(s, k) \to_H (s, 1)$ for all $s \in H$ and $k \in \{0, 1\}$.
\end{definition}

The halting states are duplicated in $\mathcal A_H$ and a self-loop is added to the copy in order to avoid introducing these stuttering steps in the middle of other executions. Since deadlock states do not have successors, this undesired situation cannot happen, so we can safely avoid duplicating deadlocked states. This construct with particular improvements will be used in~\cref{sec:mcslang}.

\subsubsection{Strategies}

	In the context of an abstract transition system $\mathcal A = (S, \to)$, strategies can be defined from different points of view~\cite{extstrat}. The following two simple and expressive characterizations will be used in this paper:
\begin{enumerate}

	\item An \emph{extensional strategy}~\cite{extensionalStrategies} is a subset $E \subseteq \Gamma_{\mathcal A}$ of the executions of $\mathcal A$.

	\item An \emph{intensional strategy} is a partial function $\lambda : S^+ \to \mathcal P(S)$ that selects the possible next steps to continue an execution $w \in S^+$ based on its history, where the states $s' \in \lambda(ws)$ must always satisfy $s \to s'$.

\end{enumerate}
Intensional strategies are less expressive than extensional strategies~\cite{extstrat}. In fact, the latter can be derived from the former by taking
$ E(\lambda) \coloneq \{ w \in S^\omega : w_{k+1} \in \lambda(w_0 \cdots w_k) \}$,
but the converse translation
$\lambda_E(w) \coloneq \{ s \in S : wsw' \in E, w' \in S^\infty \} $
loses information, and the inclusion $E \subseteq E(\lambda_E)$ could be strict. On the one hand, any $\lambda$ allows every finite prefix of an execution, while $E$ may be selective with finite traces.\footnote{In previous papers~\cite{fscd}, we extended the standard definition with an additional symbol $\lambda : S^+ \to \mathcal P(S) \cup \{\top\}$ to indicate the end of finite executions, but it causes unneeded complications.} However, for model checking, we will usually restrict to infinite traces and this is not a problem. On the other hand, the language $E(\lambda_E)$ is closed while $E$ may not be. For example, it is possible for $E$ to include the words $a^nb^\omega$ for all $n \geq 0$ but not $a^\omega$, while $\lambda_E$ will forcibly allow $a^\omega$ by definition. Nevertheless, these are properties on the infinity, that cannot be enforced by any executable and effective strategy. The study on how dropping this restriction will allow capturing fairness constraints in the strategy itself is mentioned as future work.

	In the following, the extensional notion will be mainly used for its simplicity, but usually for strategies that are also intensional, i.e.\ closed. Strategies expressed as programs in a strategy language will be translated to this abstract framework.

\subsection{Model checking} \label{sec:modelchecking}

	\emph{Model checking}~\cite{handbookmc} is a collection of automated verification techniques based on an exhaustive examination of the executions of a model to prove or refute a given property of its dynamic behavior. Models are usually described as \emph{Kripke structures} $\mathcal K = (S, \to, I, AP, \ell)$, which complement transition systems $(S, \to)$ with a set $AP$ of atomic propositions and a labeling function $\ell : S \to \mathcal P(AP)$ that declares which are satisfied on each state. It is usually assumed that the transition relation $\to$ is \emph{total}, i.e.\ that every state has a successor, to only consider infinite executions, but if it were not, we could apply the stuttering extension explained in~\cref{sec:strategies}.

	Properties are expressed in \emph{temporal logics} that extend a propositional logic whose basic predicates are atomic propositions with temporal operators to describe how they must occur in time. Logics and properties are usually divided in two classes~\cite{lamport80}:
\begin{itemize}
	\item \emph{Linear-time} properties, describing universal facts about every single execution, as if there is a unique possible future at each step. A widespread example is Linear Temporal Logic~\cite{pneuliLTL} (LTL) and its multiple extensions, but properties can also be expressed as an automaton, like the \emph{never claims} of the Spin model checker~\cite{spinmc}.
	\item \emph{Branching-time} properties refer to the whole execution tree, where multiple futures can be available at any moment in time. Well-known examples are the Computational Tree Logic~\cite{ctl} (CTL) and the more general CTL* that includes both LTL and CTL.
\end{itemize}

	The classical \emph{model-checking problem} is the question on whether a model satisfies a given property. Linear-time properties can be conveniently characterized as subsets $P \subseteq \mathcal P(AP)^\omega$, so that this is deciding whether $\ell(\Gamma^\omega_{\mathcal K}) \subseteq P$.

\subsubsection{The syntax and semantics of LTL} \label{sec:ltl-syntax}

	LTL formulae are constructed over the atomic propositions of a given Kripke structure, combined with some temporal operators.
\begin{align*}
	\varphi & \,\Coloneq\, \bot \mid \top \mid p \mid \neg\, \varphi \mid \varphi \wedge \varphi \mid \varphi \vee \varphi \mid \ctlNext \varphi \mid \ctlEvly \varphi \mid \ctlAllw \varphi \mid \varphi \ctlUntil \varphi
\end{align*}
Temporal operators describe properties of fixed execution paths: $\ctlNext \varphi$ tells that the property $\varphi$ is satisfied in the next state of the path, $\ctlEvly \varphi$ and $\ctlAllw \varphi$ say that $\varphi$ is satisfied in some or all states of the path respectively, and $\varphi_1 \ctlUntil \varphi_2$ claims that $\varphi_2$ is satisfied in some state and $\varphi_1$ holds until then. Some of these logical and temporal operators can be expressed in terms of others. 
\let\pp=\pi	The semantics of LTL formulae is usually given by a satisfaction relation $\mathcal K, \pp \vDash \varphi$ on propositional paths $\pp \in \mathcal P(AP)^\omega$:
\begin{enumerate}
		\newcommand*\ctldef[2]{\begin{tabular}{p{7em}l}#1 & iff #2\end{tabular}}
		\item \ctldef{$\mathcal K, \pp \vDash p$}{$p \in \pp_0$}
		\item \ctldef{$\mathcal K, \pp \vDash \neg\, \varphi$}{$\mathcal K, \pp \not\vDash \varphi$}
		\item \ctldef{$\mathcal K, \pp \vDash \varphi_1 \wedge \varphi_2$}{$\mathcal K, \pp \vDash \varphi_1$ and $\mathcal K, \pp \vDash \varphi_2$}
		\item \ctldef{$\mathcal K, \pp \vDash \ctlNext \varphi$}{$\mathcal K, \pp^1 \vDash \varphi$}
		\item \ctldef{$\mathcal K, \pp \vDash \varphi_1 \ctlUntil \varphi_2\,$}{$\exists \, n \geq 0 \kern1ex \mathcal K, \pp^n \vDash \varphi_2 \, \wedge \,
			\forall \, 0 \leq k < n \;\; \mathcal K, \pp^k \vDash \varphi_1$}
	\end{enumerate}
An LTL formula $\varphi$ is satisfied if $\mathcal K, \ell(\sigma) \vDash \varphi$ holds for every execution $\sigma$ of $\mathcal K$.

\subsubsection{The automata-theoretic approach for LTL model checking} \label{sec:ltl}

	\newcommand\ii{\textit{\i}}

	While other LTL model-checking algorithms exist, the explicit-state on-the-fly algorithm based on the so-called automata-theoretic approach~\cite{clarke} is probably the most widely used. This method is based on Büchi automata algorithms and the fact that the language $L(\varphi) = \{ \ell(\pi) : \mathcal K, \pi \vDash \varphi \} \subseteq \mathcal P(AP)^\omega$ of propositional traces described by an LTL formula $\varphi$ is an $\omega$-regular language~\cite{pneuliLTL}.

	The model-checking problem is equivalent to the language inclusion problem $\ell(\Gamma^\omega_{\mathcal K}) \subseteq L(\varphi)$, which is equivalent to deciding whether $\ell(\Gamma^\omega_{\mathcal K}) \cap L(\neg\,\varphi) = \emptyset$. Since $\ell(\Gamma^\omega_{\mathcal K})$ is also an $\omega$-regular language, the problem is decidable and \textsc{pspace}-complete by the results from automata theory on infinite words. Hence, model checking can be reduced to the following steps:
\begin{enumerate}
	\item Generating a Büchi automaton $B$ for $\neg \, \varphi$. The number of its states can be exponential on the size of the formula, but this is not frequent in practice.
	\item Generating an automaton $M$ for the model, usually a straightforward translation of the Kripke structure, whose transition labels are the propositional labels of the states.
	\item Calculating the intersection $L(B) \cap L(M)$, with the (synchronous) product automaton $B \times M$.
	\item Checking whether that intersection is empty, using a nested depth-first search~\cite{nestedDFS} that yields a counterexample.
\end{enumerate}
The last three steps can be performed simultaneously, generating the model automaton as required by the property, on the fly.

\subsection{Rewriting logic and Maude} \label{sec:maude}

	Rewriting logic renders change or deduction by means of rules on top of the terms of a membership equational logic~\cite{spmel}, whose \emph{signatures} are given by a set $\mathit{Sorts}$ of \emph{sorts} and a collection $\Sigma$ of operators $f : s_1 \cdots s_n \to s$ from which terms are constructed. Sorts are related by a partial order $s_1 < s_2$ representing subsort inclusion. The set of terms of a given sort $s$ over an $S$-sorted family of variables $X$ is written $T_{\Sigma, s}(X)$ and the full set of terms is written $T_{\Sigma}(X)$. Terms without variables $T_\Sigma \coloneq T_\Sigma(\emptyset)$ are called \emph{ground terms}. A \emph{substitution} is a sort-preserving function $\sigma : X \to T_\Sigma(X)$ that assigns terms to variables, and it can be extended to a function $\overline\sigma : T_\Sigma(X) \to T_\Sigma(X)$ that replaces the occurrences of the variables in a term inductively. For any pair of substitutions $\sigma_1, \sigma_2$, we define their composition $(\sigma_1 \circ \sigma_2)(x) \coloneq \overline{\sigma_2}(\sigma_1(x))$. It satisfies $\overline{\sigma_1 \circ \sigma_2} = \overline{\sigma_1} \circ \overline{\sigma_2}$ in the usual functional sense. The line over the extension is usually omitted.

	In a membership equational logic $(\Sigma, E)$, there are two classes of atomic sentences, \emph{equations} and sort \emph{membership axioms}. In their full generality, they are Horn clauses conditioned by other formulae as follows

	\[ 	t = t' \qquad \text{if } \bigwedge_i u_i = u'_i \wedge \bigwedge_j v_j : s_j \qquad\quad
		t : s \qquad \text{if } \bigwedge_i u_i = u'_i \wedge \bigwedge_j v_j : s_j \]
where $t = t'$ states that the terms $t$ and $t'$ represent the same value, $t : s$ states that $t$ has sort $s$, $u_i$, $u'_i$ and $v_j$ are arbitrary terms, $s_j$ are arbitrary sorts in the signature, and the indices $i$ and $j$ take a finite number of values. The initial algebra of all ground terms $T_\Sigma$ modulo the equality relation $=_E$ induced by the equations is written $T_{\Sigma/E}$. Its elements $[t]$ are equivalence classes, but we will usually omit the brackets and write $t$ when possible.

	Membership equational logic theories are specified in the Maude specification language as functional modules, as we will show with an example. The dining philosophers problem~\cite{csp85} is a classical concurrency problem, originally proposed by C.A.R. Hoare based on an exam exercise by E. Dijkstra. Five numbered philosophers are sat at a circular table around an endless bowl of spaghetti, and a golden fork is laid between each two contiguous philosophers. Although their main task is thinking, they should eat sometime to avoid getting starved, for what they need the two forks at both sides, which they should take one at a time and then put down when they have finished. The problem is that there are only five forks for five philosophers.
In the following functional module \texttt{PHILOSOPHERS-DINNER-BASE}, a philosopher is represented as a triple of sort \texttt{Phil} holding both hands contents of sort \texttt{Obj} (either a fork \psiCompat{} or nothing \texttt{o}) and an identifier of sort \texttt{Nat}, which is a predefined Maude sort imported from the \texttt{NAT} module. These philosophers are sat at a table of sort \texttt{Table}, which encloses a \texttt{List} of philosophers between angles.
\begin{lstlisting}[literate={psi}{\psiCompat}1, mathescape]
fmod PHILOSOPHERS-DINNER-BASE is
	protecting NAT .

	sorts Obj Phil Been List Table .
	subsorts Obj Phil < Been < List .

	ops o psi : -> Obj [ctor] .
	op (_|_|_) : Obj Nat Obj -> Phil [ctor] .
	op empty : -> List [ctor] .
	op __ : List List -> List [ctor assoc id: empty] .
	op <_> : List -> Table [ctor] .
	op initial : -> Table .

	var L : List . var P : Phil .
	eq < psi L P > = < L P psi > .
	eq initial = < (o | 0 | o) psi $\cdots$ (o | 4 | o) psi > .
endfm
\end{lstlisting}
The \texttt{ctor} attribute written next to some operator declarations indicates that they are data constructors, and the \texttt{assoc} and \texttt{id: empty} attributes for the list concatenation operator \texttt{\_\_} say that this operator is associative and that \texttt{empty} is its identity element. Structural axioms like these are treated specifically by Maude, which applies equations as reduction rules from left to right modulo these axioms, because their naive application would make the execution undecidable. The initial configuration of the problem for five philosophers is given by \texttt{initial}, where there is a fork between every two diners. Since a circular table is represented by a list, we adopt the convention that the fork between the last and first philosophers is on the right, which is ensured by the first equation.

	A \emph{rewrite theory} $\mathcal{R} = (\Sigma, E, R)$ extends the membership equational logic with a set $R$ of rewriting rules. A possibly conditional rewriting rule has the form:
\[ l \Rightarrow r \qquad \text{if } \bigwedge_i u_i = u'_i \wedge \bigwedge_j v_j : s_j \wedge \bigwedge_k w_k \Rightarrow w'_k \]
where $l$, $r$, $u_i$, $u'_i$, $v_j$, $w_k$ and $w'_k$ are any terms, and $s_j$ are any sorts for some sets of finite indices in which $i$, $j$ and $k$ range.
The application of a rule to a term $t$ is the replacement of an instance of $l$ in some position $p$ of $t$ by $r$ instantiated accordingly if the condition holds. Conditions of the third type are named \emph{rewriting conditions}, which are satisfied if the instance of each $w_k$ can be rewritten in zero or more steps to match $w_k'$.

In Maude, rewriting theories are specified in system modules where rules can be written almost as explained above. The following system module \texttt{PHILOSOPHERS\-DINNER} extends the previous equational specification of the philosophers' problem with rules so that the philosophers can take their forks and eat.
\begin{lstlisting}[literate={psi}{\psiCompat}1]
mod PHILOSOPHERS-DINNER is
	protecting PHILOSOPHERS-DINNER-BASE .

	var  Id  : Nat .    var X : Obj .    var L : List .

	rl [left]    : psi (o | Id | X) =>   (psi | Id | X) .
	rl [right]   :   (X | Id | o) psi =>   (X | Id | psi) .
	rl [left]    : < (o | Id | X) L psi > => < (psi | Id | X) L > .
	rl [release] : (psi | Id | psi) => psi (o | Id | o) psi .
endm
\end{lstlisting}
The rules \texttt{left} and \texttt{right} take the fork at the mentioned side, and \texttt{release} puts them back on the table. There is a second \texttt{left} rule for the fork between the first and last diners. Neither of the rules is conditional, but conditional rules are introduced by the \kywd{crl} keyword and are appended conditions separated by \verb|/\| after an \kywd{if} and before the dot.

	The Maude interpreter includes various commands to execute its programs~\cite{maude}. For example, \texttt{reduce} (abbreviated as \texttt{red}) simplifies a given term to its normal form with the equations and memberships $E$ modulo the structural axioms.
\begin{maudexec}[literate={psi}{\psiCompat}1]
Maude> red < psi (o | 0 | o) > .
rewrites: 1
result Table: < (o | 0 | o) psi >
\end{maudexec}
The \texttt{rewrite} (\texttt{rew}) command rewrites a term using all the rewriting rules in the module, until a normal form is found or up to an optional number of rewriting steps given between brackets.
\begin{maudexec}[literate={psi}{\psiCompat}1]
Maude> rew [4] initial .
rewrites: 12
result Table: < (psi | 0 | o) (psi | 1 | psi) (o | 2 | o)
                (psi | 3 | o) psi (o | 4 | o) >
\end{maudexec}
Moreover, the \texttt{search} command lets the user find all terms reachable by rewriting that match a pattern and satisfy a specified condition. The rewriting paths that lead to the found terms can also be inspected. For example, we can check the presence of deadlock states on the dining philosophers problem using a search for normal forms \texttt{=>!}.
\begin{maudexec}[literate={psi}{\psiCompat}1]
Maude> search initial =>! T:Table .

Solution 1 (state 211)
states: 243  rewrites: 932
T:Table --> < (psi | 0 | o) (psi | 1 | o) (psi | 2 | o)
              (psi | 3 | o) (psi | 4 | o) >

Solution 2 (state 242)
states: 243  rewrites: 980
T:Table --> < (o | 0 | psi) (o | 1 | psi) (o | 2 | psi)
              (o | 3 | psi) (o | 4 | psi) >

No more solutions.
states: 243  rewrites: 980
\end{maudexec}
The command shows two states where each fork is taken by a different philosopher, so that no one can take the other one and eat, causing the starvation of the whole group. This problem will be solved using strategies in the following sections. More details about the language and the interpreter can be found in the Maude manual~\cite{maude}.

	Rewriting logic and Maude specifications can be seen as transition systems $(T_{\Sigma/E},$ $\to^1_R)$ whose states are terms and whose transitions are one-step rule rewrites. Temporal properties can be checked on this model using the Maude LTL model checker~\cite{maudemc}, which is an integral part of Maude since its 2.0 version and has been given many applications. Our model checker for strategy-controlled systems is an extension of this tool, and they are used in a very similar way. Consequently, the details on how Maude specifications are prepared for model checking and the decidability conditions in~\cref{sec:maudesmc} are a close adaptation of what the Maude manual explains for the standard one~\cite{maude}. The builtin model checker is an optimized implementation of the standard on-the-fly LTL algorithm described in~\cref{sec:ltl} using the LTL2BA algorithm~\cite{fastLTL} with some optimizations~\cite{efficientBuchi}.

\section{Model checking abstract strategy-controlled systems} \label{sec:smc}

	Understanding the satisfaction of temporal properties on systems controlled by strategies is clearer when they are seen in the abstract and generic terms of~\cref{sec:strategies} rather than as syntactic expressions on a strategy language. Given a strategy-controlled system $(\mathcal K, E)$, the main intuition is that temporal properties should be checked on the executions allowed by the strategy $E$, regardless of the others. This motivates the following natural definition for linear-time properties.
\begin{definition} \label{def:smc}
	Given a strategy-controlled system $(\mathcal K, E)$ and a linear-time property $\varphi$, $(\mathcal K, E) \vDash \varphi$ if $\mathcal K, \ell(\pi) \vDash \varphi$ for all $\pi \in E$.
\end{definition}

	Remember that linear-time properties are universally satisfied by every execution of a model, and a satisfaction relation on propositional traces is always well-defined. Branching-time properties can be contemplated similarly, since strategies also restrict the branches of the execution trees where they are checked. These properties are addressed in~\cite{btimemc}, so in this paper we will focus on linear-time ones.

	In order to use this definition with concrete strategy descriptions, like expressions in strategy languages, we should indicate which executions are allowed by them. This is done for the Maude strategy language by means of a small-step operational semantics in~\cref{sec:slang}. However, some relevant consequences of the previous abstract definition are valid in general:
\begin{itemize}
	\item The satisfaction of a temporal property solely depends on the executions allowed by the strategy, and not on its concrete representation.
	\item Any temporal logic or property that is well defined in the base system is also well defined when it is controlled by a strategy.
	\item Conversely, the properties under consideration do not reason about strategies, but about the system that results from their restrictions.
\end{itemize}

	Considering the language $L(\varphi) \coloneq \{ \rho \in \mathcal P(AP)^\omega : \mathcal K, \rho \vDash \varphi \}$ of propositional traces admitted by $\varphi$, the model-checking problem is reduced to a language inclusion $\ell(E) \subseteq L(\varphi)$, whose decidability, complexity and algorithmic results can be exploited. If the property logic is LTL, $L(\varphi)$ is an $\omega$-regular language and the problem is \textsc{pspace}-complete for any $\omega$-regular strategy $E$, and \textsc{2exptime}-complete for any $\omega$-context-free strategy~\cite{pushdownLTLComplexity}, but the program complexity (for a fixed formula) in both cases is polynomial on the size of the automaton.
Moreover, if $E$ is $\omega$-regular the automata-theoretic approach explained in~\cref{sec:ltl} can be applied even if the automaton for $\ell(E)$ has non-trivial Büchi conditions.\footnote{In the automata-theoretic approach (see~\cref{sec:modelchecking}), the intersection of the model automaton $L(\mathcal K)$ and the negated property automaton $L(\neg \varphi)$ is calculated to decide $\mathcal K \vDash \varphi$. In this case, the automaton for $L(\mathcal K)$ has trivial Büchi conditions and the intersection algorithm is simpler. However, if $L(\mathcal K)$ is replaced by an $\ell(E)$ with non-trivial Büchi conditions, a similar intersection algorithm can be applied, although the required space may double~\cite{handbookmc}.} More precisely, it is $\ell(E)$ that has to be $\omega$-regular or $\omega$-context-free, but the same properties on $E$ are sufficient conditions.

	In order to model check strategy-controlled systems with off-the-shelf algorithms for the appropriate logics, finding a Kripke structure whose executions coincide with $E$ or only whose propositional traces coincide with $\ell(E)$ is a general approach. How to build this structure or how to transform $\mathcal K$ accordingly may be specific for each strategy language or formalism. For the Maude strategy language, this structure will be generated using the small-step operational semantics in the following section. In general, it is certain that such a Kripke structure exists and it is finite iff $\ell(E)$ is a closed and $\omega$-regular language.

\begin{proposition}
	Given $E \subseteq S^\omega$, there is a finite Kripke structure $\mathcal K'$ such that $\ell(\Gamma^\omega_{\mathcal K'}) = \ell(E)$ iff $\ell(E)$ is closed and $\omega$-regular.
\end{proposition}
For branching-time properties, the coincidence of the propositional traces is not enough and a stronger bisimulation relation is required~\cite{btimemc}.

\section{The Maude strategy language and its semantics} \label{sec:slang} \label{sec:opsem}

	The Maude strategy language~\cite[\S 10]{maude} controls rewriting on Maude specifications. Its most basic component is the selective application of rules, which are combined with typical programming constructs to describe complex rewriting strategies. Its syntax is summarized in the following grammar from the $\alpha$ symbol:
{\renewcommand\skywd[1]{\texttt{#1}}
\begin{align*}
	\alpha & \,\Coloneq\, \beta \mid \texttt{top(}\beta\texttt{)} \mid \idle \mid \fail \mid \cond PC \mid \alpha \seq \alpha \mid (\alpha \disj \alpha) \mid \alpha \,\texttt* \mid \ifthel\alpha\alpha\alpha \\
		& \;\;\mid\;\;\, \texttt{matchrew} \;P\; \texttt{s.t} \;C \; \texttt{by} \; x \;\texttt{using} \; \alpha , \; \ldots , x \; \texttt{using} \; \alpha \mid \mathit{slabel} \mid \mathit{slabel}\texttt(\vec t\texttt) \\
		& \;\;\mid\;\;\, \alpha \,\texttt+ \mid \alpha \,\texttt! \mid \alpha \;\texttt{or-else}\; \alpha \mid \texttt{test(}\alpha\texttt{)} \mid \texttt{try(}\alpha\texttt{)} \mid \texttt{not(}\alpha\texttt{)} \\
	\beta	& \,\Coloneq\, \mathit{rlabel} \mid \mathit{rlabel}\texttt[\rho\texttt] \mid \mathit{rlabel}\texttt\{\vec\alpha\texttt\} \mid \mathit{rlabel}\texttt[\rho\texttt]\texttt\{\vec\alpha\texttt\} \mid \texttt{all} \\
	\rho	& \,\Coloneq\, x \;\texttt{<-}\; t \mid  x \;\texttt{<-}\; t \;\texttt,\; \rho
\end{align*}}The core of the language is in the first two rows, including the rule application strategies in the $\beta$ symbol, since strategy combinators in the third row can be defined in terms of those of the first two. The semantics of strategy expressions is usually described by the terms that result of rewriting from an initial term under its control~\cite{strategies06}. This is what the Maude command \texttt{srewrite $t$ using $\alpha$} and its depth-first variant \texttt{dsrewrite} show when evaluating strategies. However, in order to check temporal properties on rewriting systems controlled by this language, explicitly stating the intermediate states of computations is essential, as we will do with a small-step operational semantics.\footnote{Another rewriting-based operational semantics had been proposed before for the language~\cite{rewSemantics}. However, tracing the rewriting sequence of a term out of the executions of this semantics is more complicated than with the semantics used in this paper.} This semantics will connect the strategy language with the previous section and its abstract definition of model checking for strategies.

	Since the evolution of the rewriting process with a strategy depends at any time both on the current term and execution state of the strategy, the semantics is defined on a set of augmented states $\xs$ univocally associated to a term by a projection $\cterm : \xs \to T_\Sigma$. Augmented states are essentially pairs of a term and a strategy continuation $t \ao \alpha_1 \cdots \alpha_n$ where the ordered execution of $\alpha_1$ to $\alpha_n$ is pending from $t$, but richer structure is required to support strategy calls and the execution of some complex combinators of the language. Their syntax is defined by the $q$ symbol of the following grammar
\begin{align*}
	q & \,\Coloneq\, t \ao s \mid \subterm(x: q, \ldots, x: q; t) \ao s \mid \rewcond(x: q, \theta, C, \vec\alpha, \theta, t, t; t) \ao s \\
	s & \,\Coloneq\, \varepsilon \mid \alpha s \mid \theta s \\
	\vec\alpha & \,\Coloneq\, \alpha \mid \alpha \vec\alpha
\end{align*}
where the terminal symbol $t$ stands for terms, $x$ for variables, $\alpha$ for strategy expressions, $\theta$ for substitutions, and $C$ for rule conditions. The non-terminal $s$ represents stacks of pending strategy expressions and substitutions. Substitutions will be pushed in certain situations like strategy calls, and the active substitution for a given stack $s$ will be written $\vctx(s)$ and determines the values of the variables in the strategy expressions. This function can be defined recursively as $\vctx(\theta s) = \theta$, $\vctx(\varepsilon) = \mathrm{id}$, and $\vctx(\alpha s) = \vctx(s)$. States with an empty stack $t \ao \varepsilon$ have nothing pending and are called \emph{solutions}. The \emph{current term} $\cterm$ projection can also be defined structurally, with $\cterm(t \ao s) = t$ being its base case. The constructors of the strategy language are the following:
\begin{itemize}
	\item Rule applications, indicating the label $\mathit{rlabel}$ of the rule and some optional restrictions. \[ t \ao \mathit{rlabel}\texttt[x_1 \,\texttt{<-}\, t_1\texttt, \ldots\texttt, x_n \,\texttt{<-}\, t_n] \; s \to_s t' \ao s \]
The optional substitution $\rho$ that maps $x_i$ to $t_i$ between brackets is applied to both sides of the rule and its condition before matching, in order to restrict its application or allow applying rules with free variables.\footnote{Maude allows declaring rules with free variables in its righthand side and condition, but they must be marked with the \texttt{nonexec} attribute, and can only be further used at the metalevel after instantiation or for narrowing~\cite[\S\ 4.5.3]{maude}.} A rule with $m$ rewriting conditions can be executed if exactly $m$ strategies are provided between brackets to control their evaluation. In the small-step semantics, this is specified using the $\rewcond$ state,
\begin{align*}
	 t &\ao \mathit{rl}\hbox{\tt[}x_1 \,\hbox{\tt<-}\, t_1, \ldots, x_n \,\hbox{\tt<-}\, t_n\hbox{\tt]}\hbox{\tt\{}\alpha_1, \ldots, \alpha_k\hbox{\tt\}} \, s \\
	&\to_c \rewcond(p_1 : \sigma(l_1) \ao \alpha_1 \theta, \sigma, C', \alpha_2 \cdots \alpha_k, \theta, r, c ; t) \ao s
\end{align*}
In this execution state, a subsearch is started from the lefthand side of every rewriting condition fragment $l_k \,\texttt{=>}\, p_k$ of the selected rule from left to right. These lefthand sides are instantiated with the substitution $\sigma$ carried by the $\rewcond$ state and determined by the initial substitution, and the evaluation of the previous equational and rewriting condition fragments. However, the variables in the strategy expression and the initial substitution are given value by the environment $\theta = \vctx(s)$. When a solution is found for a rewriting fragment and it matches the righthand side pattern $p$, the evaluation continues with the next one after updating the substitution $\sigma'$ accordingly.
\[ \begin{array}{l}
	\rewcond(p: t' \ao \varepsilon, \sigma, C_0 \wedge l \ttrew p' \wedge C, \alpha \vec \alpha, \theta, r, c ; t) \ao s \\[2pt]
	\kern1em\to_c \rewcond(p': \sigma'(l) \ao \alpha \, \theta, \sigma', C, \vec \alpha, \theta, r, c ; t) \ao s
\end{array} \]
When the last fragment is solved, the term is finally rewritten by putting the righthand side of the selected rule $r$ instantiated by the accumulated substitution $\sigma'$ in the context $c$ where the lefthand side of the rule matched.
\[ \rewcond(p: t' \ao \varepsilon, \sigma, C_0, \vec \alpha, r, c ; t) \ao s \to_s c(\sigma'(r)) \ao s \]
Notice that the very first and last rules execute a \emph{system transition} $\to_s$ while the others take a \emph{control transition} $\to_c$, since the former are applying a rule in the underlying rewriting systems while the latter only do some auxiliary strategic work. This distinction will be useful when extracting rewriting paths from executions of the semantics. To conclude with the $\rewcond$ search, the substate included in the execution state is another execution state that is executed similarly.

\unskip
	\begin{prooftree}
		\AxiomC{$q \to_\bullet q'$}
		\UnaryInfC{$\rewcond(p: q, \sigma, C, \vec \alpha, \theta, r, c ; t) \ao s \to_c \rewcond(p: q', \sigma, C, \vec \alpha, \theta, r, c ; t) \ao s$}
	\end{prooftree}
However, both control and system transitions on the inner state are control transitions on the outer one, since it is an auxiliary term and not the subject term what is being rewritten. Thus, the state includes a copy of the initial term, so that we can define $\cterm(\rewcond(\ldots; t)) = t$.

Rules are applied anywhere by default, but matching can be limited to the topmost position by surrounding the strategy with \skywd{top}. Another special rule application operator is \skywd{all}, which executes any rule in the module with the usual behavior.
	\item Tests \lstinline[mathescape]{match $P$ s.t. $C$} check whether the subject term matches the pattern $P$ and satisfy the equational condition $C$.
\[ t \ao (\cond{P}{C}) \, s \to_c t \ao s \qquad \text{if $t$ matches $P$ and satisfies $C$} \]
The test is simply popped when it succeeds, and the execution gets blocked otherwise. The initial keyword can be changed to \skywd{amatch} to match anywhere, or to \skywd{xmatch} to match with extension (see~\cite[\S\ 4.8]{maude}).
	\item Strategies can be combined with a series of operators like concatenation $\alpha \seq \beta$ that executes $\beta$ on the results produced by $\alpha$.
\[ t \ao (\alpha \seq \beta) \, s \to_c t \ao \alpha \beta \, s \]
In the semantics, they are pushed to the stack of pending strategies in that order. The union $\alpha \disj \beta$ executes $\alpha$ or $\beta$ nondeterministically.
\[ t \ao (\alpha \disj \beta) \, s \to_c t \ao \alpha \, s \qquad t \ao (\alpha \disj \beta) \, s \to_c t \ao \beta \, s \]
And the iteration $\alpha \texttt*$ repeatedly executes $\alpha$ a nondeterministic number of times.
\[ t \ao \alpha^* \, s \to_c t \ao s \qquad t \ao \alpha^* \, s \to_c t \ao \alpha \alpha^* \, s \]
Together with the constants $\idle$ and $\fail$, which do nothing and interrupt the execution respectively, this family of combinators resembles those of regular expressions.
\[ t \ao \idle \, s \to_c t \ao s \qquad \text{no rule for \fail} \]
There is no rule for $\fail$, so it blocks the execution like a failed test. In general, we say that a strategy \emph{fails} if it does not produce any result.

	\item The conditional operator \lstinline[mathescape]{$\alpha$ ? $\beta$ : $\gamma$} executes its condition $\alpha$ first. If it does not fail, its solutions are continued by the positive branch $\beta$. Otherwise, $\gamma$ is executed from the initial term.
\[ t \ao (\ifthel\alpha\beta\gamma) \, s \to_c t \ao \alpha \beta \, s \]
The previous rule can always be applied, since $\beta$ will not be executed if $\alpha$ fails. The negative branch is only executed when $\alpha$ has been evaluated exhaustively without finding solutions, where ${\to_{s,c}} = {\to_s} \cup {\to_c}$.

	\begin{prooftree}
		\AxiomC{$\to_{s,c}$ is terminating from $t \ao \alpha \, \theta$ and does not reach solutions}
		\LeftLabel{[else]}
		\UnaryInfC{$t \ao \ifthel\alpha\beta\gamma \, s \to_c t \ao \gamma \, s$}
	\end{prooftree}

	\item The combinator \lstinline[mathescape]{matchrew $P$ s.t. $C$ by $x_1$ using $\alpha_1$, $\ldots$, $x_n$ using $\alpha_n$} allows rewriting selected subterms of the subject term.
\begin{align*}
	t \ao &(\matchrewf) \, s \\
	 &\to_c \subterm(x_1 : \sigma(x_1) \ao \alpha_1 \, \sigma , \ldots , x_n : \sigma(x_n) \ao \alpha_n \, \sigma ; \sigma_{-\{x_1, \ldots, x_n\}}(P)) \ao s
\end{align*}
The subterms matching the variables $x_k$ in the pattern $P$ are rewritten according to the corresponding strategies $\alpha_k$ in parallel. This pattern and the condition are previously instantiated with the variable context $\vctx(s)$, and substrategies may also use their variables in addition to the environment ones.
\begin{prooftree}
	\AxiomC{$q_i \to_\bullet q_i'$}
	\UnaryInfC{$\subterm(\ldots, x_i: q_i, \ldots ; t) \ao s \to_\bullet \subterm(\ldots, x_i: q_i', \ldots ; t) \ao s$}
\end{prooftree}
The multiple execution states are executed concurrently with both control and system transitions. This is well defined because rewriting naturally occurs inside subterms, and so we define
\[ \cterm(\subterm(x_1 : q_1, \ldots, x_n : q_n, t) = t[x_1/\cterm(q_1), \ldots, x_n/\cterm(q_n)].\]
Finally, when solutions have been found for all the subterms, the original term is reassembled with them in place of the original subterms.
\[ \subterm(x_1: t_1 \ao{} \varepsilon, \ldots, x_n: t_n \ao{} \varepsilon ; t) \ao s \to_c t[x_1 / t_1, \ldots, x_n / t_n] \ao s \]
There are \skywd{amatchrew} and \skywd{xmatchrew} variants like for tests. Another interesting usage of this operator is obtaining information about the subject term by instantiating variables to be used in the strategy control logic, as shown in the examples of~\cref{sec:examples}.
	\item Finally, it is possible to give name to strategy expressions and define them in strategy modules, as we will explain soon. These named strategies are called by writing their names followed by a comma-separated list of arguments between parentheses, if any.
\[ t \ao \mathit{sl}(t_1, \ldots, t_n) \, s \to_c t \ao \delta \, \sigma \, s \]
All the definitions $\delta$ in the module whose lefthand side matches with $\sigma$ the call term will be executed nondeterministically. The call term is instantiated with $\vctx(s)$ before matching, and its substitution is popped $t \ao \sigma \, s \to_c t \ao s$ when the strategy call finishes. In case of tail calls, when the top of the stack $s$ is a substitution, this substitution can be replaced by the new one following the typical optimizations of programming languages.
Recursive and mutually recursive definitions are allowed, thus increasing the expressive power of the language. 
\end{itemize}
Some more combinators can be derived from these like \lstinline[mathescape]{$\alpha$ or-else $\beta$} defined as \lstinline[mathescape]{$\alpha$ ? idle : $\beta$}, \lstinline[mathescape]{not($\alpha$)} as \lstinline[mathescape]{$\alpha$ ? idle : fail}, \lstinline[mathescape]{try($\alpha$)} as \lstinline[mathescape]{$\alpha$ ? idle : idle},  \lstinline[mathescape]{test($\alpha$)} as \lstinline[mathescape]{not(not($\alpha$))}, and the normalization operator $\alpha \texttt!$ as \lstinline[mathescape]{$\alpha$ * ; not($\alpha$)}.

	The nondeterministic small-step operational semantics given in the previous paragraphs by the $\to_s$ and $\to_c$ transitions defines which are the rewriting paths allowed by any strategy expression $\alpha$, i.e., its extensional denotation as in~\cref{sec:strategies}. Looking at these rules, we can conclude that system steps $q \to_s q'$ correspond to rule rewrites $\cterm(q) \to^1_R q'$ on the underlying module, and control steps $q \to_c q'$ do not alter the subject term $\cterm(q) = \cterm(q')$. Hence, we define the relation ${\opsem} \coloneq {\to_c^*} \circ \to_s$ that executes a system step preceded by as many control steps as required, as the basis for extracting rewriting paths out of semantic executions.

\begin{definition} \label{def:mslstrat}
	Given a strategy expression $\alpha$ and a term $t \in T_\Sigma$, we define
\[ E(\alpha) \coloneq \cup_{t \in T_\Sigma} E(\alpha, t) \qquad E(\alpha, t) \coloneq \cterm(\mathrm{Ex}^*(\alpha, t) \cup \mathrm{Ex}^\omega(\alpha, t)) \]
where $\mathrm{Ex}^*(\alpha, t) \coloneq \{ q_0 q_1 \cdots q_n : q_0 = t \ao \alpha, q_k \opsem q_{k+1}, q_n \to_c^* t' \ao \varepsilon, t' \in T_\Sigma \}$ and $\mathrm{Ex}^\omega(\alpha, t) \coloneq \{ (q_k)_{k=0}^\infty : q_0 = t \ao \alpha, q_k \opsem q_{k+1} \}$.
\end{definition}

The elements of $E(\alpha)$ are clearly rewriting paths, where each term is connected with the next one by a rule rewrite, since they are the projection of executions of the semantics by the $\opsem$ relation. $\mathrm{Ex}^\omega(\alpha, t)$ is the set of all nonterminating executions of $\alpha$, and $\mathrm{Ex}^*(\alpha, t)$ contains all complete finite executions, those ending in a state where a solution can be reached by control steps.
This semantics does not only provide the abstract strategy definition, but also engenders a labeled transition system on which to model check using standard algorithms as suggested in the previous sections. Moreover, this transition system inspires the actual implementation of the model checker for strategy-controlled systems, presented in~\cref{sec:maudesmc,sec:implementation}. 

	Recovering the example of the dining philosophers, we can execute the following strategy to make a philosopher take its forks using the \texttt{srewrite} command, which shows the last states of the finite executions in $E(\alpha)$, or in other words, the solutions $t \ao \varepsilon$ reachable by $\to_{s,c}$ steps.
\begin{maudexec}[literate={psi}{\psiCompat}1]
Maude> srewrite psi (o | 0 | o) psi using left ; right .

Solution 1
rewrites: 2
result Table: (psi | 0 | psi)

No more solutions.
rewrites: 2
\end{maudexec}
What we do not see with \texttt{srewrite} is the whole rewriting path and its intermediate steps.
\begin{center}
\begin{tikzpicture}[every node/.style={anchor=west}, y=2.5em, x=.6\linewidth]
	\node (S0) at (0, 0) {\texttt{\psiCompat{} (o | 0 | o) \psiCompat} \ao\; \texttt{left ; right}};
	\node (S1) at (0, -1) {\texttt{\psiCompat{} (o | 0 | o) \psiCompat} \ao\; \texttt{left} \, \texttt{right}};
	\node (S2) at (0, -2) {\texttt{\hphantom{\psiCompat} (\psiCompat{} | 0 | o) \psiCompat} \ao\; \texttt{right}};
	\node (S3) at (0, -3) {\texttt{\hphantom{\psiCompat} (\psiCompat{} | 0 | \psiCompat)} \ao\; $\varepsilon$};

	\node[right=1ex of S0] (T0) at (1, 0) {\texttt{\psiCompat{} (o | 0 | o) \psiCompat}};
	\node[right=1ex of S2] (T2) at (1, -2) {\texttt{\hphantom{\psiCompat} (\psiCompat{} | 0 | o) \psiCompat}};
	\node[right=1ex of S3] (T3) at (1, -3) {\texttt{\hphantom{\psiCompat} (\psiCompat{} | 0 | \psiCompat) \hphantom{\psiCompat}}};

	\draw[->] (.415, -.3) -- node {\scriptsize $c$} (.415, -.7);
	\draw[->] (.415, -1.3) -- node {\scriptsize $s$} (.415, -1.7);
	\draw[->] (.415, -2.3) -- node {\scriptsize $s$} (.415, -2.7);
	\draw[->] (T0) -- node[very near end] {\scriptsize $R$} node[left, very near end] {\scriptsize $1$} (T2);
	\draw[->] (T2) -- node {\scriptsize $R$} node[left] {\scriptsize $1$} (T3);
\end{tikzpicture}
\end{center}
We can also obtain the 243 reachable states of the problem with five philosophers by rewriting \texttt{initial} with \texttt{(left | right | release)} \texttt{*} among others.

\subsection{Strategy modules}

	For more elaborate strategies, strategy modules are useful to give them names and define them compositionally. These modules extend functional and system modules with strategy declarations and definitions. They start by the \texttt{smod} keyword and end with \texttt{endsm}. Strategies are declared with a name and signature, including the sort of the parameters it may receive $s_1, \ldots, s_n$, and the sort $s$ of the term to which it will be applied, although the latter is only informative.
\begin{lstlisting}[mathescape]
strat $\mathit{sname}$ : $s_1$ $\ldots$ $s_n$ @ $s$ .
\end{lstlisting}
Multiple strategies with a common signature can be defined at once writing multiple names, and in this case the plural keyword \lstinline{strats} is preferred. Strategies are defined with statements similar to equations and rules, but whose righthand side is a strategy expression that may contain free variables occurring in the lefthand side strategy call or in the equational condition $C$.
\begin{lstlisting}[mathescape]
sd $\mathit{sname}$($p_1$, $\ldots$, $p_n$) := $\alpha$ .
csd $\mathit{sname}$($p_1$, $\ldots$, $p_n$) := $\alpha$ if $C$ .
\end{lstlisting}
These definitions may contain strategy calls so that recursive and mutually recursive strategies can be defined. As already mentioned, all matching strategy definitions are executed on a strategy call. Deeper explanations about the strategy language can be found in~\cite[\S\ 10]{maude}.

	Coming back to the running example, we will extend it with a strategy module. The uncontrolled execution of this system is not satisfactory for the philosophers integrity, as we have seen with the \texttt{search} command and as we will see soon by model checking, so some restrictions are specified using strategies. These are gathered in a strategy module \texttt{DINNER-STRAT} that extends and controls \texttt{PHILOSOPHERS-DINNER}.
\begin{lstlisting}
smod DINNER-STRAT is
	protecting PHILOSOPHERS-DINNER .

	strats free parity turns @ Table .
	strat turns : Nat Nat @ Table .

	var  T    : Table .
	vars L L' : List .
	vars K Id : Nat .
	var  N    : NzNat .
	vars X Y  : Obj .
\end{lstlisting}
The first strategy, \texttt{free}, is the recursive and exhaustive application of all the rules in the module, and so it behaves like the builtin strategy of the \texttt{rewrite} command.
\begin{lstlisting}
	sd free := all ? free : idle .
\end{lstlisting}
Assuming that the philosophers in the table are numbered consecutively from zero, the equivalent of the solution proposed by Dijkstra to solve the original exam exercise is the \texttt{parity} strategy. It forces the diners to take first the fork at a fixed side, which is alternative for even and odd, i.e., for neighbors.
This restriction groups the philosophers in pairs where they compete for the middle fork, and only the one with this fork will try to obtain the outer fork shared with another couple, hence not impeding their other neighbors to take both forks and eat.
\begin{lstlisting}[literate={psi}{\psiCompat}1{:=}{{:=}}2{<-}{{<-}}2]
	sd parity := (release
		*** The even take the left fork first
		| (amatchrew L s.t. psi (o | Id | o) := L
		     /\ 2 divides Id by L using left)
		| left[Id <- 0]
		*** The odd take the right fork first
		| (amatchrew L s.t. (o | Id | o) psi := L
		     /\ not (2 divides Id) by L using right)
		*** When they already have one, they take the other fork
		| (amatchrew L s.t. (psi | Id | o) psi := L
		     by L using right)
		| (matchrew T s.t. < L (o | Id | psi) L' > := T
		     by T using left[Id <- Id])
		) ? parity : idle .
\end{lstlisting}
The last strategy, \texttt{turns}, iterates through the philosophers in a loop, making them eat in turns. The strategy can be improved by allowing more than one philosopher to eat in parallel (with five philosophers, two can eat at each turn).
\begin{lstlisting}[literate={psi}{\psiCompat}1{:=}{{:=}}2{<-}{{<-}}2]
	sd turns(K, N) := left[Id <- K] ; right[Id <- K] ;
	                  release ; turns(s(K) rem N, N) .
	sd turns := matchrew T s.t. < L (X | Id | Y) psi > := T
	             by T using turns(0, s(Id)) .
endsm
\end{lstlisting}
The argument \texttt{N} of the first \texttt{turns} strategy is the number of philosophers at the table, and \texttt{K} is the cyclic index to the current one. Their initial values are filled by the overloaded version without arguments, which obtains the number of philosophers from the initial term.

	In \cref{sec:philosophers}, some temporal properties will be checked on this strategy-controlled model. For the moment, we can anticipate that the \texttt{parity} strategy solves the deadlock problem of the uncontrolled system. Since the recursion of \texttt{parity} stops when no rule can be applied, after jumping to the negative branch of the conditional, the \texttt{srewrite} command would show a deadlock state if it existed, but it does not.
\begin{maudexec}
Maude> srew initial using parity .

No solution.
rewrites: 709
\end{maudexec}

\newcommand\Sol{\mathrm{Sol}}

\section{Model checking for the Maude strategy language} \label{sec:mcslang}

After discussing the meaning of model checking for strategy-controlled systems in \cref{sec:smc} and describing the rewriting paths allowed by an expression in the Maude strategy language in \cref{sec:slang}, the satisfaction of linear-time properties in Maude specifications controlled by its strategy language is already unambiguously defined.

	Suppose we are given a rewrite theory $\mathcal R = (\Sigma, E, R)$ specified in a Maude module $M$, and an additional signature $\Pi$ of atomic propositions defined on the terms of $\mathcal R$ by some equations $D$ using a satisfaction predicate \texttt{\_|=\_}. The Kripke structure of the uncontrolled rewriting is defined as
\[ \mathcal M \coloneq (T_{\Sigma/E}, \to^1_R, T_{\Sigma/E}, AP_\Pi, L_\Pi) \]
where $\to^1_R$ is the one-step rewrite relation,
\[ AP_\Pi \coloneq \{ \; \theta(p(x_1, \ldots, x_n)) \mid p \in \Pi, \theta \text{ ground substitution} \; \} \]
is the set of ground instances of the atomic proposition terms, and
\[ L_\Pi([t]) \coloneq \{ \theta(p(x_1, \ldots, x_n)) \in AP_\Pi \mid (E \cup D) \vdash \; t \vDash \theta(p(x_1, \ldots, x_n)) = \texttt{true} \} \]
is the labeling function that evaluates them under the equations $E$ and $D$. Given a strategy expression $\alpha$ in $M$, possibly referring to some strategy definitions in the module, and a linear-time property $\varphi$ on the previous atomic propositions, $\varphi$ is satisfied in $M$ controlled by $\alpha$ if
\[ (\mathcal M, E(\alpha)) \vDash \varphi \iff \forall \pi \in E(\alpha) \quad \mathcal M, L_\Pi(\pi) \vDash \varphi \]
according to \cref{def:smc,def:mslstrat}. The extensional strategy $E(\alpha)$ may contain finite traces where logics like LTL are not properly defined, but these can be extended to infinite ones by the typical stuttering extension explained in~\cref{sec:strategies}.

	In order to reuse existing model-checking algorithms for the target logic, a general approach is proposed in \cref{sec:smc}, finding a Kripke structure whose propositional traces coincide with $L_\Pi(E(\alpha))$. A reasonable candidate is the graph of the nondeterministic small-step operational semantics of~\cref{sec:slang}, $\ltssl^{\alpha, t} \coloneq (\xs, \opsem, \{ t \ao \alpha \}, AP_\Pi, \linebreak \cterm \circ L_\Pi)$.\footnote{Assuming that the Kripke structure has a single initial state instead of finitely many is without loss of generality, since each initial state can be treated separately.} Indeed, the nonterminating executions of $\ltssl^{\alpha, t}$ projected by the $\cterm$ function are the nonterminating rewriting paths of $E(\alpha, t)$ by definition of $\mathrm{Ex}^\omega(\alpha, t)$ and $E(\alpha, t)$. Finite executions pose more problems since they should be extended to infinite traces, but only those that are complete executions of the strategy $\alpha$. As defined in $\mathrm{Ex}^*(\alpha, t)$, these are the executions ending in states $\Sol \coloneq \{ q \in \xs : q \to_c^* \cterm(q) \ao \varepsilon  \}$ where a solution can be reached by control steps. Using the construction of \cref{def:halting} with $H = \Sol$, the Kripke structure that represents $E(\alpha, t)$ can be defined as
\[ \ltmsl \coloneq \ltssl^{\alpha, t}_\Sol = (\xs \times \{0\} \cup \Sol \times \{1\}, \opsem_\Sol, \{ t \ao \alpha \}, AP_\Pi, \pi_1 \circ \cterm \circ L_\Pi) \]
where $\pi_1$ is the projection of the first component. The transition system $\ltmsl$ implements the stuttering extension on the finite traces of $\ltssl^{\alpha, t}$ by adding a self-loop to solution states, where finite executions are allowed to terminate.

\begin{proposition} \label{prop:osol}
	The projection of the infinite traces of $\ltmsl$ by $\pi_1 \circ \cterm$ coincides with the stuttering-extension of $E(\alpha, t)$.
\end{proposition}

\begin{figure}[h]\centering
\begin{tikzpicture}[state/.style={}]
	\node[state] (S1) {\small\texttt{fail}};
	\node[state, right=8em of S1] (S2) {\small $t' \ao \varepsilon$};
	\node[state, right=8em of S2] (S3) {\small $q$};
	\node[state, above right=-.4em and 1.7em of S3] (S31) {\small $q'$};
	\node[state, dashed, below right=-.4em and 1.7em of S3] (S32) {\small $t' \ao \varepsilon$};

	\draw[->>] (S3) -- (S31);
	\draw[->>, dashed] (S3) -- (S32);

	\draw (S2) edge[->, dashed, loop right] ();
	\draw (S32) edge[->, dashed, loop right] ();

	\draw[->>] (S1.west) + (-2ex, 0) -- (S1.west);
	\draw[->>] (S2.west) + (-2ex, 0) -- (S2.west);
	\draw[->>] (S3.west) + (-2ex, 0) -- (S3.west);

\node[below=of S1] (L1) {\small (1) Dead end};
	\node[right=2em of L1] (L2) {\small (2) Deadlock solution state};
	\node[right=2ex of L2] {\small (3) Continuable solution state};
\end{tikzpicture}
\caption{Solution and deadlock states in $\ltssl^{\alpha, t}$ and their adjustments.} \label{fig:deadlock}
\end{figure}
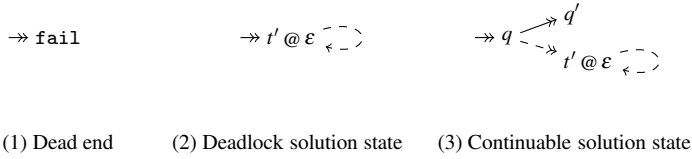

	However, the abstract construction of $\ltssl^{\alpha, t}_\Sol$ can be applied more efficiently in this particular case. In effect, there are three relevant situations regarding finite traces, shown in \cref{fig:deadlock}. In the third case, where the solution state has a successor that allows continuing the execution, its duplication is justified. This situation may occur for example after executing $\beta$ in the strategy $\beta \texttt*$, when both finishing the iteration and continuing with $\beta$ are possible. If the loop were added directly to the solution state, spurious executions would be allowed that stay a number of steps in the solution state and then continue by its successor. This situation cannot happen in the second case, where the solution state does not have successors, so a loop can be safely added to it without duplication.

	In the first case, the state is not a solution, but one in which the strategy has failed. Since no loop is added to it and only the nonterminating executions of $\ltmsl$ are considered, this execution state is completely ignored, as well as all other states from which neither solution states nor infinite executions can be reached. From the point of view of the strategy, these states and the executions that go through them have been discarded by an explicit \skywd{fail}, a failed test, an inapplicable rule, etc., and so they are seen as if they have never happened. These failed states do not disturb the standard on-the-fly LTL algorithm described in \cref{sec:ltl} because its nested depth-first search will not find any cycle through them. Removing failed states can be done in linear-time complexity on the number of states by an exploration of the rewriting graph similar to the Tarjan's SCC algorithm~\cite{tarjan}, but this is incompatible with on-the-fly model checking because the entire graph might need be explored to conclude that a single state is valid. However, this removal algorithm must be surely applied for other model-checking algorithms that do not enjoy this property, like tableau-based ones for LTL.

	In conclusion, the rewriting system controlled by $\alpha$ can be checked against linear-time properties with the standard algorithms for the desired logic using the just defined $\ltmsl$.

\begin{corollary}
	$(\mathcal M, E(\alpha, t)) \vDash \varphi \iff \ltmsl \vDash \varphi$ for any linear-time property $\varphi$. 
\end{corollary}

	Model checking against the most usual temporal logics is decidable whenever the Kripke structure is finite, and its transition relation and labeling function are computable. In the case of $\ltmsl$, this does not only depend on the strategy and the finiteness of its execution space for the operational semantics, but also on other requirements of the rewriting specification shared with the standard model checker~\cite[\S 12.3]{maude}. These latter conditions are the typical executability requirements for Maude modules ensuring that applying rules on normal forms as Maude does is lossless. Given a strategy-controlled specification as specified before, model checking is well defined and decidable if:
\begin{itemize}
	\item The rewrite theory $\mathcal R = (\Sigma, E, \phi, R)$ specified by $M$ plus the equations $D$ defining the predicates $\Pi$ satisfy:
	\begin{itemize}
		\item both $E$ and $E \cup D$ are (ground) Church-Rosser and terminating perhaps modulo axioms, where $(\Sigma, E) \subseteq (\Sigma \cup \Pi, E \cup D)$ is a protecting extension, i.e.\ it adds neither junk nor confusion to what it extends,
		\item $R$ is (ground) coherent relative to $E$ perhaps modulo axioms.
	\end{itemize}
	\item The set of reachable execution states from $t \ao \alpha$ by $\to_{s,c}$ is finite (this implies $\to_{s,c}$ and $\opsem$ are decidable, see~\cref{lem:decidable}). This set can be defined as $\{ q : t \ao \alpha \to_{s,c}^* q \}$.
\end{itemize}

We are not specifying the linear-time logic in which properties will be expressed, although only Linear Time Logic is actually used in this paper. Of course, these decidability conditions may be excessive if the logic is trivial enough, and insufficient if it is extremely complex.

The finiteness of the set of reachable states is related with other aspects of the abstract strategy and the strategy expression. How strategy-controlled Maude specifications are checked in practice is discussed in \cref{sec:maudesmc}.

\subsection{Expressiveness of the language and decidability} \label{sec:express}

	In this section, we briefly discuss what can be specified and model checked using the Maude strategy language. While the language is Turing complete, only strategies whose denotations lie in more restricted language classes will make model checking decidable with the proposed method. Since strategies describe subsets of executions of a fixed system, the Turing completeness of a strategy language can be understood as the ability to denote any recursive enumerable subset of traces of a given transition system. This property is trivially met by the Maude strategy language having stateful recursive definitions.

\begin{proposition} \label{prop:turingcomplete}
	For any $\infty$-recursively enumerable language $L \subseteq \Gamma_{\mathcal M}$, there is some strategy expression $\alpha$ such that $E(\alpha) = L$.
\end{proposition}

	In the previous section, we have concluded that model checking is decidable for LTL and similar logics on well-behaved rewriting systems iff the reachable states of the operational semantics are finitely many. In that case, the language of rewriting paths denoted by the strategy expression is $\infty$-regular.

\begin{proposition} \label{prop:fin-closedregular}
	If the reachable states from $t \ao \alpha$ are finitely many, $E(\alpha, t)$ is a closed $\infty$-regular language.
\end{proposition}

	The converse of~\cref{prop:fin-closedregular} is not true, as the strategy expression \texttt{empty(0)} with the definition \texttt{empty(N) := fail | empty(s(N))} clearly shows. The language denoted by \texttt{empty(0)} is the $\omega$-regular and closed empty set, but infinitely many execution states are reachable when \texttt{empty} is called with increasing arguments. However, this example is very artificial and an alternative strategy like $\fail$ can be found to describe the same language. This is possible in general, since the strategy language includes regular expressions as a sublanguage.

\begin{proposition} \label{prop:ext2expr}
	If $L$ is a closed $\infty$-regular language, there is a strategy expression $\beta$ such that $E(\beta) = L$ and the reachable states from $t \ao \beta$ are finitely many for all $t \in T_\Sigma$.
\end{proposition}

	However, the iteration is not a faithful representation of the Kleene star, because it allows executing its body infinitely many times. This is why only closed languages can be described with Maude strategies.

	In addition to the previous conditions based on the language properties of the strategy, other conditions can be formulated in terms of syntactic properties of the expressions. Strategies are potentially complex recursive programs depending on the rewriting system and equational theory in which they are applied, so simple conditions can only be obtained for very particular cases. Strategies without recursive calls or iterations always produce finitely many states, but its usefulness is very limited. Assuming that only a finite number of terms are involved in the execution of the strategy, iteration and tail-recursive strategies can be called while keeping the state space finite, even if the calls do not terminate. Remember that a tail call is a call executed as the last action of the calling context, which can be located at the syntactical end of expressions.

\begin{definition}
	All recursive calls are tail in a strategy expression if it is:
\begin{itemize}
	\item \idle, \fail, a test, or a strategy call expression.
	\item $\alpha \disj \beta$ if all recursive calls in $\alpha$ and $\beta$ are tail.
	\item $\alpha \seq \beta$ if $\alpha$ does not contain recursive calls and all recursive calls in $\beta$ are tail.
	\item $\ifthel\alpha\beta\gamma$ if $\alpha$ does not contain recursive calls, and all recursive calls in $\beta$ and $\gamma$ are tail.
	\item A subterm rewriting or rule application expression, if all recursive calls in its substrategies are tail.
\end{itemize}
\end{definition}

\begin{definition} \label{def:reachableterms}
	The set of reachable terms from $t \ao \alpha$ is $\bigcup_{q \in \{ q : t \lower1pt\hbox{\scriptsize @} \alpha \to_{s,c}^* q\}} \mathrm{terms}(q)$ where
\[ \mathrm{terms}(q) \coloneq \cterm(q) \cup \begin{cases}
		\cup_k \mathrm{terms}(q_k)	& \text{if } q = \subterm(\ldots, x_k : q_k, \ldots) \\
		\mathrm{terms}(q')						& \text{if } q = \rewcond(x : q', \ldots) \\
		\{ \theta(t_1), \ldots, \theta(t_n) \} 			& \text{if } q = t \ao \mathit{sl}\,(t_1, \ldots, t_n) \, s
	\end{cases} \]
and where $\theta = \vctx(s)$.
\end{definition}

\begin{proposition} \label{prop:finiterec}
	The reachable states from $t \ao \alpha$ are finitely many if any of the following conditions holds:
	\begin{enumerate}
		\item $\alpha$ does not contain iterations or recursive calls.
		\item The reachable terms from $t \ao \alpha$ are finitely many and all recursive calls in $\alpha$ and the reachable strategy definitions are tail.
	\end{enumerate}
\end{proposition}

The number of reachable states can be explicitly bounded in terms of the length and other syntactical properties of the strategy expression and the number of states of the uncontrolled model. However, that bound will not be satisfactory in most cases since the strategy and the visited terms are completely dependent of each other. In any case, the states of $\mathcal \ltmsl$ are fewer, since they are calculated with the $\opsem$ relation. 

\section{The Maude strategy-aware model checker} \label{sec:maudesmc} \label{sec:ltlmc}

	We have extended the builtin Maude LTL model checker~\cite{maudemc} to support rewriting systems controlled by its strategy language, based on the foundations of the previous sections. The original LTL model checker implements the automata-theoretic approach (explained in~\cref{sec:ltl}) clearly separating its three components: the generation of a Büchi automaton for the temporal property, the on-the-fly generation of an automaton for the model, and the algorithm that checks whether the intersection of the previous two is empty. Strategies only restrict the executions of the model and do not interfere with the property specification, so only the second of these parts has been adapted by replacing the standard rewrite system on $(T_\Sigma, \to^1_R)$ with the strategy-aware model $\ltmsl$ described in~\cref{sec:mcslang}, with the help of the infrastructure for executing strategies of the \texttt{srewrite} and \texttt{dsrewrite} commands. Consequently, a significant part of the C++ and Maude implementation of the model checker has been reused, and the interfaces of both model checkers are very similar, so that users of the original can use the strategy-aware one without much effort. Actually, they can be used simultaneously on the same module to compare the properties of the controlled and uncontrolled system.
The extension is not exempt of subtleties and difficulties that are described in more detail in~\cref{sec:implementation}.

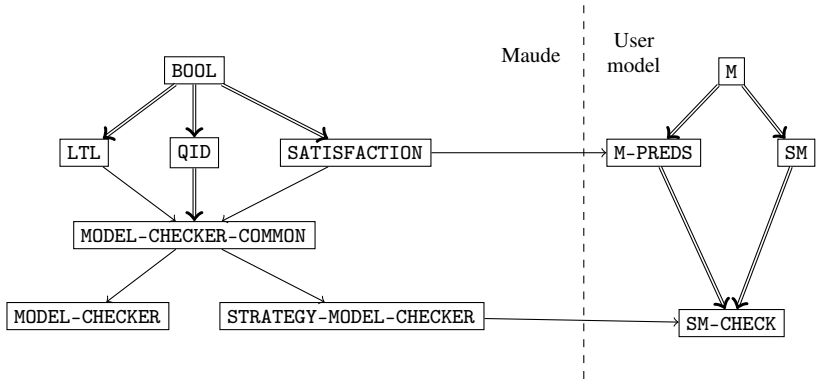
\begin{figure}[t]\centering
	\begin{tikzpicture}[node distance=.7cm and .8cm]
		\node[draw] (SAT) {\ttfamily SATISFACTION};
		\node[draw, left=of SAT] (QID) {\ttfamily QID};
		\node[draw, left=of QID] (LTL) {\ttfamily LTL};

		\node[draw, above=of QID] (BOOL) {\ttfamily BOOL};

		\node[draw, below=of QID] (CC) {\ttfamily MODEL-CHECKER-COMMON};

		\node[draw, below left=.7cm and -4.5em of CC] (C) {\ttfamily MODEL-CHECKER};
		\node[draw, below right=.7cm and -4.5em of CC] (SC) {\ttfamily STRATEGY-MODEL-CHECKER};

		\node[draw, right=2.3cmof SAT] (MP) {\ttfamily M-PREDS};
		\node[right=.3cm of MP] (anchor) {\vphantom{M}};
		\node[draw, right=.5cm of anchor](SM) {\ttfamily SM};
		\node[draw, above=of anchor](M) {\ttfamily M};
		\node[draw, below=1.88cm of anchor] (MC) {\ttfamily SM-CHECK};

		\draw[->] (SAT) -- (MP);
		\draw[double, ->] (M) -- (MP);
		\draw[double, ->] (M) -- (SM);
		\draw[double, ->] (MP) -- (MC);
		\draw[double, ->] (SM) -- (MC);
		\draw[double, ->] (QID) -- (CC);
		\draw[->] (LTL) -- (CC);
		\draw[->] (SAT) -- (CC);
		\draw[->] (CC) -- (C);
		\draw[->] (CC) -- (SC);
		\draw[->] (SC) -- (MC);
		\draw[double, ->] (BOOL) -- (LTL);
		\draw[double, ->] (BOOL) -- (QID);
		\draw[double, ->] (BOOL) -- (SAT);

\draw[dashed] (3cm, -3cm) -- +(0, 5cm);
		\node[anchor=east] at (2.8cm, 1.3cm) {\small Maude};
		\node[anchor=west] at (3cm, 1.3cm) {\small \begin{tabular}cUser\\model\end{tabular}};
	\end{tikzpicture}
	\caption{Structure of the strategy model checker modules.} \label{fig:smcheck}
\end{figure}

	\cref{fig:smcheck} outlines how strategy-aware models are typically prepared for model checking, where some modules available in the Maude prelude or provided by the model checker in its \texttt{model-checker.maude} file are involved. The process is done in much the same way as explained in the Maude manual~\cite{maude} for the original model checker. The input model is given by a system module \texttt{M} describing the uncontrolled system and a strategy module \texttt{SM} defining one or more strategies to control its behavior.\footnote{The separation of the modules \texttt{M} and \texttt{SM} in the model specification is a matter of style. In general, we propose specifying the static model representation and the rules in a system module \texttt{M}, and describing how they are controlled in a strategy module \texttt{SM} being a protecting extension of \texttt{M}.} In order to specify atomic propositions, a module \texttt{M-PREDS} is defined as a protecting extension of \texttt{M} where the builtin \texttt{SATISFACTION} module is included, providing the formal sorts \texttt{State} for the model states and \texttt{Prop} for atomic propositions, and the symbol \texttt{\_|=\_} to define with equations whether these atomic propositions are satisfied in each state.
\begin{lstlisting}
fmod SATISFACTION is
	protecting BOOL .
	sorts State Prop .
	op _|=_ : State Prop -> Bool [frozen] .
endfm
\end{lstlisting}
The intended state sort in \texttt{M} is defined as a subsort of \texttt{State} in \texttt{M-PREDS}, along with the declaration and definition of the atomic propositions, i.e., the signature $\Pi$ and the equations $D$ mentioned in \cref{sec:mcslang}. Finally, both \texttt{M-PREDS} and \texttt{SM} are gathered in a strategy module \texttt{SM-CHECK} that includes the module \texttt{STRATEGY-MODEL-CHECKER}. This module is the entry point to the model checker via a special \texttt{modelCheck} operator that receives the problem data and reduces to its verification result:
\begin{lstlisting}
op modelCheck : State Formula Qid QidList Bool
                 -> ModelCheckerResult [special(...)] .
\end{lstlisting}
The first and second arguments coincide with the \texttt{modelCheck} operator of the classical model checker: the initial term and the LTL formula to be checked, with the syntax specified in the \texttt{LTL} module and using the atomic propositions in \texttt{M-PREDS}.
\begin{lstlisting}
ops True False : -> Formula [ctor ...] .
op ~_ : Formula -> Formula [ctor prec 53 ...] .
op _/\_ : Formula Formula -> Formula [comm ctor prec 55 ...] .
op _\/_ : Formula Formula -> Formula [comm ctor prec 59 ...] .
op O_ : Formula -> Formula [ctor prec 53 ...] .
op _U_ : Formula Formula -> Formula [ctor prec 63 ...] .
op _->_ : Formula Formula -> Formula [prec 65 ...] .
op <>_ : Formula -> Formula [prec 53 ...] .
op []_ : Formula -> Formula [prec 53 ...] .
\end{lstlisting}
The name of the strategy without arguments that must control the system is specified in its third argument as a quoted identifier. Alternatively, an arbitrary strategy expression can be provided using the unified Maude model-checking tool \texttt{umaudemc}~\cite{umaudemc,btimemc}, which has a graphical and command-line interface.
\begin{quotation}
\newcommand*\nonterm[1]{$\langle\,$\textit{#1}$\,\rangle$}
\newcommand*\terminal[1]{\texttt{#1}}
\terminal{umaudemc} \terminal{check} \nonterm{file name} \nonterm{initial term} \nonterm{formula} \nonterm{strategy}
\end{quotation}
The \texttt{modelCheck} symbol and this tool incorporate two other optional arguments that allow considering the execution of some strategies as atomic transitions and enable a particular form of partial order reduction regarding \texttt{matchrew} combinators, which are explained in~\cref{sec:opaque,sec:biased}. Formally, reducing the term \texttt{modelCheck($t$, $\varphi$, '$\mathit{sname}$)} triggers the model checking of $(\mathcal M, E(\mathit{sname}, t)) \vDash \varphi$.
If the property is satisfied, the operator is reduced to the constant \texttt{true} of sort \texttt{Bool}. In case the property does not hold, the returned counterexample \texttt{counterexample($\pi$, $\xi$)} consists of a path $\pi$ and a cycle $\xi$ such that $\pi\xi^\omega \in E(\mathit{sname}, t)$ is an execution allowed by the strategy that refutes the formula. The syntax of counterexamples does not differ much from those of the standard model checker, the sequences are a juxtaposition of transitions \texttt{\{$t$, $r$\}} where $t$ is a term and $r$ describes the transition that rewrites this term into the next one by the name of the rule that has been applied or the constant \texttt{unlabeled} if it does not have one. In the last transition of the cycle, an $r$ can take the value \texttt{solution} to indicate that a finite strategy execution refutes the given property. These explanations are illustrated with the dining philosophers and other examples in~\cref{sec:examples}.

	When using the alternative \texttt{umaudemc} interface, the counterexample is instead shown in the terminal or displayed as a graph. This tool also allows obtaining graphs of the strategy-controlled transition system used internally by the model checker, and checking linear-time properties expressed in other logics like the $\mu$-calculus~\cite{btimemc}.

	The fourth and fifth arguments of the \texttt{modelCheck} operator, and the equivalent options of the \texttt{umaudemc} tool, allow deviating intentionally from the model specified by the semantics in~\cref{sec:opsem}. We explain them in the following sections.

\subsection{Opaque strategies} \label{sec:opaque}

	The main principle of our understanding of model checking for strategy-controlled systems is that its executions are a subset of those of the original system. In addition to its theoretical convenience, the principle has practical implications when model checking linear-time properties, since refuting a property on a system controlled by a strategy refutes the property for the original system. However, it can be sometimes useful to deviate from this rule and consider the execution of some strategies as atomic steps, rather than the rule rewrites they consist of. Such a sequence of several $\opsem$ steps can be seen as a single step, with transitions linking the state where the strategy has been called to those in which its execution concludes.

	Strategies whose executions are considered atomic are called \emph{opaque strategies} and passed to the model checker as a list of strategy names in its fourth argument. The list cannot discriminate between homonym strategies with a different signature, but renaming the desired strategy using the Maude renaming support is easier that admitting signature specifications there. In the \texttt{modelCheck} result, opaque strategies appear as \texttt{opaque($s$)} in place of the rule name where $s$ is the name of the strategy. 

	Strategies are suitable to represent parallel rewriting or specify systems in which the rule rewrites are not the meaningful steps of the model. In these cases, opaque strategies can be used to make these steps agree with the semantics of the system. Moreover, opaque strategies can also be used to test properties at different levels of granularity. For instance, the computational model of membrane systems can be represented in rewriting logic~\cite{membraneJournal} implementing its evolution steps by the execution of multiple rule rewrites controlled by a strategy. The whole strategy execution and not each rule application is the atomic step in this model, so opaque strategies can be used to contemplate them as single transitions when model checking~\cite{memstratmc}. They are also used in the example of \cref{sec:opsemimp}.

\subsection{Biased matchrew as a form of partial order reduction} \label{sec:biased}

The intended meaning of the \texttt{matchrew} family of combinators is the parallel rewriting of the matched subterms using the specified strategies. However, executions are seen as linear sequences of rule rewrites, so the rewrites coming from the different subterms must be ordered. The small-step semantics permits the progress of any subterm at any moment, hence considering all possible interleavings of the subterm rewriting paths as executions of the \texttt{matchrew}. This is semantically accurate but computationally expensive, since even in the case of only two subterms with a single rewriting path allowed for each of them, this yields the binomial coefficient $n + m$ over $n$ of interleaved executions where $n$ and $m$ are the length of these paths.
When the model checker users know that the ordering of the subterm paths does not affect the satisfaction of the property in question, they can choose to exhibit only one representative to the model checker as a form of partial order reduction, avoiding the generation of the full set of combinations. The $n + m$ over $n$ executions of the small hypothetical case above are reduced to a single one. This is specified in the optional fifth Boolean argument of the \texttt{modelCheck} operator, where \texttt{true} indicates that the \emph{biased matchrew} feature is enabled, the option used by default. Specifically, the biased executions have the rewrites ordered like the subterms in the \texttt{matchrew} term from left to right, so that all the rewrites of the $k$ subterm occur before those of the $k+1$.

	For example, we can informally consider a system with two processes and a shared resource, and the following \texttt{matchrew} as part of the strategy that controls the system:
\begin{lstlisting}
matchrew < P1, P2, SR > by P1 using step ! ,
                           P2 using step ! .
\end{lstlisting}
Supposing that this strategy advances the processes until they need the shared resource, and the property refers only to the shared resource ownership, the property will be satisfied or refuted regardless of the interleaving of the processes states. However, if the property refers to certain relationships between the two processes, ignoring some executions may miss counterexamples that refute the property.

\section{Examples} \label{sec:examples}

	The strategy-aware model checker has been applied to various examples~\cite{stratweb}, including classical concurrency algorithms, reactive systems, process algebras, telecommunication protocols, other computational models~\cite{memstratmc}, games~\cite{metatrans}, and so on. The model checker being publicly available for some time, it has already been independently used to model check properties of smart contracts~\cite{bitmlmc}.

	The main goal of the first of the three examples included in this section is to introduce the model checker and the procedure to have specifications model checked by it. In the second one, strategies are used to represent alternative scheduling policies in a multiprocessor and the model checker is used to determine whether some properties like fairness are satisfied depending on them. The third example is a paradigmatic use case of strategies, a small-step operational semantics including negative premises and rules with different priorities, whose programs are checked by our tool.

\subsection{The philosophers problem} \label{sec:philosophers}

	In this section, we will illustrate how to model check a strategy-controlled system with the dining philosophers example. Remember that the signature and rules of the problem have been specified in the system module \texttt{PHILOSOPHERS-DINNER} in \cref{sec:maude}, and some strategies have been defined in the strategy module \texttt{PHILOSOPHERS-STRAT} in \cref{sec:slang}. Following the procedure described in~\cref{fig:smcheck}, this system module is \texttt{M}, the complete specification of the uncontrolled model. The extension \texttt{DINNER-PREDS} below specifies the atomic propositions that will be used to describe properties of the problem behavior: a parameterized collection \texttt{eats($n$)} meaning ``the philosopher $n$ eats'', and \texttt{used($n$)} standing for ``the fork at the right of philosopher $n$ is being used''.
\begin{lstlisting}[literate={psi}{\psiCompat}1]
mod DINNER-PREDS is
	protecting PHILOSOPHERS-DINNER .
	including SATISFACTION .

	subsort Table < State .
	ops eats used : Nat -> Prop [ctor] .

	var  Id  : Nat .
	var  X   : Obj .
	vars L R : List .

	eq < L (psi | Id | psi) R > |= eats(Id) = true .
	eq < L > |= eats(Id) = false [owise] .

	eq < L (X | Id | o) psi R > |= used(Id) = false .
	eq < L > |= used(Id) = true [owise] .
endm
\end{lstlisting}
Notice that \texttt{Table} is declared as a subsort of \texttt{State}, and equations are used to define the satisfaction of the atomic propositions on every state. 

	Finally, the strategy specification in \texttt{DINNER-STRAT} is merged with the property specification in \texttt{DINNER-PREDS}. In the same module, an \texttt{initial} operator is defined to build the initial table with the given number of philosophers, which is five by default \texttt{< (o | 0 | o) \psiCompat{} $\cdots$ (o | 4 | o) \psiCompat{} >}. The rules and strategies of the model are valid regardless of the number of philosophers, which is determined by the initial term.
\begin{lstlisting}[literate={psi}{\psiCompat}1]
smod DINNER-SCHECK is
	protecting DINNER-STRAT .
	protecting DINNER-PREDS .

	op initial     :     -> Table .
	op initial     : Nat -> Table .
	op initialList : Nat -> List .

	eq initial = initial(5) .
	eq initial(N) = < initialList(N) > .
	eq initialList(0) = empty .
	eq initialList(s(N)) = initialList(N) (o | N | o) psi .
endsm
\end{lstlisting}

	Now, we can start model checking. The property that would guarantee the survival of the philosophers is the LTL property $\ctlAllw \bigwedge_{k=0}^4 \ctlEvly \texttt{eats($k$)}$, but the unrestricted system does not even satisfy the weaker non-deadlock requirement $\ctlAllw \ctlEvly \bigvee_{k=0}^4 \texttt{eats($k$)}$.
\begin{maudexec}[literate={psi}{\psiCompat}1, escapechar=^]
Maude> red modelCheck(initial,
             <> (eats(0) \/ ^\ldots^ \/ eats(4)), 'free) .
rewrites: 120
result ModelCheckResult: counterexample(
	{< (o |0| o) psi (o |1| o) psi (o |2| o) psi
	   (o |3| o) psi (o |4| o) psi >,'left}
	{< (psi |0| o) psi (o |1| o) psi (o |2| o) psi
	   (o |3| o) psi (o |4| o) >,'left}
	{< (psi |0| o) (psi |1| o) psi (o |2| o) psi
	   (o |3| o) psi (o |4| o) >,'left}
	{< (psi |0| o) (psi |1| o) (psi |2| o) psi
	   (o |3| o) psi (o |4| o) >,'left}
	{< (psi |0| o) (psi |1| o) (psi |2| o)
	   (psi |3| o) psi (o |4| o) >,'left},
	{< (psi |0| o) (psi |1| o) (psi |2| o)
	   (psi |3| o) (psi |4| o) >,solution})
\end{maudexec}
In this counterexample, every philosopher takes the left fork before anyone can take the right one and eat. While the system has been checked using the \texttt{free} strategy, using the standard model checker is equivalent and the same counterexample is obtained, although with \texttt{deadlock} instead of \texttt{solution} as the last transition label. Instead, deadlock is avoided with the \texttt{parity} strategy:
\begin{maudexec}[literate={psi}{\psiCompat}1, escapechar=^]
Maude> red modelCheck(initial, 
            [] <> (eats(0) \/ ^\ldots^ \/ eats(4)), 'parity) .
rewrites: 1005
result Bool: true
\end{maudexec}
However, it does not ensure that no philosopher starves.
\begin{maudexec}[literate={psi}{\psiCompat}1, escapechar=^]
Maude> red modelCheck(initial,
            [] (<> eats(0) /\^ \ldots ^/\ <> eats(4)), 'parity) .
rewrites: 558
result ModelCheckResult: counterexample(
	{< (o |0| o) psi (o |1| o) psi (o |2| o) psi
	   (o |3| o) psi (o |4| o) psi >,'left}
	{< (o |0| o) psi (o |1| o) (psi |2| o) psi
	   (o |3| o) psi (o |4| o) psi >,'left}
	{< (o |0| o) psi (o |1| o) (psi |2| o) psi
	   (o |3| o) (psi |4| o) psi >,'left}
	{< (psi |0| o) psi (o |1| o) (psi |2| o) psi
	   (o |3| o) (psi |4| o) >,'right}
	{< (psi |0| o) psi (o |1| o) (psi |2| psi)
	   (o |3| o) (psi |4| o) >,'release},
	{< (psi |0| o) psi (o |1| o) psi (o |2| o) psi
	   (o |3| o) (psi |4| o) >,'right}
	{< (psi |0| o) psi (o |1| psi) (o |2| o) psi
	   (o |3| o) (psi |4| o) >,'left}
	{< (psi |0| o) (psi |1| psi) (o |2| o) psi
	   (o |3| o) (psi |4| o) >,'release})
\end{maudexec}
Not all problems of the counterexample above can be attributed to conflicts between philosophers. The only philosopher eating repeatedly in this trace is 1, but 3 and 4 could have eaten on their own, since they do not share any fork with 1. In fact, the strategy does not require the philosophers to eat whenever possible, although it can be modified to enforce it. Alternatively, a premise can be added to the LTL property $\ctlAllw \bigwedge_{k=1}^4 \ctlEvly \texttt{used($k$)}$ to ensure that no fork is underused. Anyhow, this does not prevent starvation.
\begin{maudexec}[escapechar=^]
Maude> red modelCheck(initial,
        [] (<> used(0) /\ ^\ldots^ /\ <> used(4))
     -> [] (<> eats(0) /\ ^\ldots^ /\ <> eats(4)), 'parity) .
rewrites: 4455
result ModelCheckResult: counterexample(^\ldots^, ^\ldots^)
\end{maudexec}
The omitted counterexample consists of eleven steps and shows that 0 and 3 do not eat because 1 and 2 are always \emph{faster} to take their shared fork. In order to avoid starvation completely an external synchronization source is required~\cite{csp85}. For example, a simple but perhaps too forced solution is establishing turns as in the \texttt{turns} strategy.
\begin{maudexec}[escapechar=^]
Maude> red modelCheck(initial,
        [] (<> eats(0) /\ ^\ldots^ /\ <> eats(4)), 'turns) .
rewrites: 541
result Bool: true
\end{maudexec}
Despite the nonterminating recursive definition of \texttt{turns}, the model checker terminates thanks to its ability to detect cycling tail-recursive calls even with arguments.

\subsection{Processes and scheduling policies} \label{sec:scheduling}

	The computers that we use in our everyday life are continuously running multiple interactive processes that share their resources~\cite{modernOS}. Even if the number of physical and logical processors included in modern chips grows endlessly, the list of simultaneous processes increases too and the operating system has to decide which processes are granted access to the processing units at each time so that all tasks get done without unnecessary delay and degradation of the user experience. Moreover, these processes may depend on and communicate with each other and with external peripherals. Scheduling policies are strategies of the operating system to arrange the computer execution time, and in this example we will represent very simple instances of these in the Maude strategy language and check how properties are satisfied depending on them.

	The simplified computer model used in this section is based on the Maude implementation of the Dekker algorithm in~\cite{allmaude,maudemc}. It consists of a shared memory composed of integer cells indexed by the name of the variables, and a soup of processes running in the same processor.
\begin{lstlisting}
sort Memory .
op [_,_] : Qid Int -> Memory [ctor] .
op none  : -> Memory [ctor] .
op __    : Memory Memory -> Memory
             [ctor assoc comm id: none] .

sorts Pid Process Soup MachineState .
subsort Process < Soup . subsort Int < Pid .
op [_,_]   : Pid Program -> Process [ctor] .
op empty   : -> Soup [ctor] .
op _|_     : Soup Soup -> Soup
               [ctor prec 61 assoc comm id: empty] .
op {_,_,_} : Soup Memory Pid -> MachineState [ctor] .
\end{lstlisting}
The third component of the machine state is the identifier of the last process that has been run of sort \texttt{Pid}, which includes the integers as a subtype. Processes consist of a process identifier and a program in a simple imperative programming language:
\begin{lstlisting}
sorts Test UserStatement Program .
subsort UserStatement < Program .
ops skip io       : -> Program [ctor] .
op _;_            : Program Program -> Program 
                      [ctor prec 61 assoc id: skip] .
op _:=_           : Qid Int -> Program [ctor] .
op _=_            : Qid Int -> Test [ctor] .
op if_then_fi     : Test Program -> Program [ctor] .
op while_do_od    : Test Program -> Program [ctor] .
op repeat_forever : Program -> Program [ctor] .
\end{lstlisting}
The language constructs and their meaning are standard, where \lstinline{;} is sequential composition and \lstinline{:=} is assignment. Their semantics are defined by means of rules that manipulate the machine state. For instance, the rule for \texttt{repeat} is
\begin{lstlisting}
vars I J : Pid .             var M : Memory .
vars P R : Program .         var S : Soup .

rl [exec] : {[I, repeat P forever ; R] | S, M, J} 
         => {[I, P ; repeat P forever ; R] | S, M, I} .
\end{lstlisting}
The sort \texttt{UserStatement} may include other statements that are consumed when encountered, and the \texttt{io} instruction executes some input/output operation that is treated differently in the following.\begin{lstlisting}
var U : UserStatement .

rl [exec] : {[I, U ; R] | S, M, J} => {[I, R] | S, M, I} .
rl [io]   : {[I, io ; R] | S, M, J} => {[I, R] | S, M, I} .
\end{lstlisting}
Semaphores are also supported in the language with their two operations \texttt{wait} and \texttt{signal} implemented by the following rules:
\begin{lstlisting}
ops wait signal : Qid -> Program [ctor] .

var Q : Qid .                var N : Int .

crl [exec] : {[I, wait(Q) ; R] | S, [Q, N] M, J} 
          => {[I, R] | S, [Q, N - 1] M, I} if N > 0 .
rl  [exec] : {[I, signal(Q) ; R] | S, [Q, N] M, J} 
          => {[I, R] | S, [Q, N + 1] M, J} .
\end{lstlisting}
The rule for \texttt{wait} fails if the memory value \texttt{N} in \texttt{Q} is not greater that zero, and so processes in that situation will not advance.

	Using this language the following programs can be written: they execute a critical section (\texttt{crit} is defined as a user statement) protected by a semaphore in the \texttt{mutex} variable.
\begin{center}
\begin{tabular}{c@{\hspace{.05\linewidth}}c}
\begin{lstlisting}
eq p = repeat
          wait('mutex) ;
          crit ;
          signal('mutex)
       forever .
\end{lstlisting}
&
\begin{lstlisting}
eq pIo = repeat
            wait('mutex) ;
            crit ;
            signal('mutex) ;
            io
         forever .
\end{lstlisting}
\end{tabular}
\end{center}
The \texttt{pIo} program additionally executes an input/output operation outside the critical section.

	In the rewriting system described above, the \texttt{exec} rule tries to run any process in the soup nondeterministically, and so their execution is completely concurrent. Even so, semaphores are enough to guarantee that only one process is in the critical section at the same time. To check this, we define an atomic proposition \texttt{inCrit($k$)} that tells whether the process $k$ is in the critical section, extending as usual the system module.
\begin{lstlisting}
subsort MachineState < State .
eq {[I, crit ; R] | S, M, J} |= inCrit(I) = true .
eq MS |= inCrit(I) = false [owise] .
\end{lstlisting}
Moreover, since the property claiming that no pair of processes are simultaneously in the critical section
\[ \textstyle \ctlAllw \neg \, \left( \bigvee_{n=0}^N \bigvee_{m=0}^{n-1} \texttt{inCrit($n$)} \wedge \texttt{inCrit($m$)} \right) \]
is verbose and depends on the number $N$ of processors, we built it equationally together with the initial configuration.
\begin{lstlisting}
op onlyOne       : Nat         -> Formula .
op inCritFormula : Nat Nat     -> Formula .
op initial       : Nat Program -> MachineState .

vars N M : Nat .             var P : Program .
eq onlyOne(N) = []~ inCritFormula(N, N) .
eq inCritFormula(0, 0) = False .
eq inCritFormula(1, s(M)) = inCritFormula(M, M) .
eq inCritFormula(s(N), M) = (inCrit(N) /\ inCrit(M)) 
     \/ inCritFormula(N, M) [owise] .

eq initial(N, P) = { initialSoup(N, P), ['mutex, 1], 0 } .
eq initialSoup(0, P) = empty .
eq initialSoup(s(N), P) = initialSoup(N, P) | [s(N), P] .
\end{lstlisting}
The mutual exclusion property specified above can be checked with the standard model checker for any fixed number of processors.
\begin{maudexec}
Maude> red modelCheck(initial(4, p), onlyOne(4)) .
rewrites: 4373
result Bool: true
\end{maudexec}
However, it is not true that every process eventually gets into the critical section (we omit the counterexample because it has 39 states).
\begin{maudexec}[mathescape]
Maude> red modelCheck(initial(4, p), <> inCrit(1)) .
rewrites: 201
result ModelCheckerResult: counterexample($\ldots$, $\ldots$)
\end{maudexec}

On top of this specification and in a separate strategy module, we have defined different scheduling policies as strategies. Since changing the active process involves the expensive operation of saving or restoring its execution context, operating systems try to amortize it by executing as many instructions as possible before changing again. The \texttt{blocked} policy keeps executing the current process with \lstinline{exec[I <- P]} where the last process \texttt{P} has been obtained with the \skywd{matchrew} from the machine state. However, if this process is blocked by an \texttt{io} operation or in a semaphore, the rule \texttt{exec} executes any other process nondeterministically.
\begin{lstlisting}
sd blocked := ((matchrew MS s.t. {S, M, P} := MS
                by MS using exec[I <- P])
                        or-else (try(io) ; exec))
              ; blocked .
\end{lstlisting}
Another common scheduling policy is called \emph{round-robin}. The \texttt{roundRobin} strategy maintains in its argument a list of process identifiers and tries to execute them cyclically, passing to the next state when the current one gets blocked. The process list can be initially empty or incomplete, in case it is filled nondeterministically with the available processes.
\begin{lstlisting}[escapechar=^]
sd roundRobin(nil) := matchrew MS
				s.t. {[P, R] | S, M, J} := MS
				by MS using (exec[I <- P] ; roundRobin(P)) .
sd roundRobin(P LP) := exec[I <- P] ? roundRobin(P LP) : (
	try(io) ;
	((matchrew MS s.t. {[I, R] | S, M, J} := MS
	    /\ ^not^(occurs(I, P LP)) by MS using exec[I <- I])
	? (matchrew MS s.t. {S, M, I} := MS
	    by MS using roundRobin(I LP P))
	: roundRobin(LP P))
\end{lstlisting}
However, a process can still occupy the processor forever. The round-robin policy can be modified to be \emph{preemptive} by assigning a maximum time slice for each process and pass the usage of the processor to the next one once it is consumed, if it was not blocked before.

\begin{lstlisting}[escapechar=^]
sd roundRobin(P LP, 0, N) := try(io) ; (
	(matchrew MS s.t. {[I, R] | S, M, J} := MS
	   /\ ^not^(occurs(I, P LP)) by MS using exec[I <- I])
	? (matchrew MS s.t. {S, M, I} := MS
	     by MS using roundRobin(I LP P, N, N))
	: roundRobin(LP P, N, N)) .

sd roundRobin(P LP, s(K), N) := exec[I <- P] ?
     roundRobin(P LP, K, N) : roundRobin(P LP, 0, N) .
\end{lstlisting}

	Since the uncontrolled model already protects the critical section, and because all linear-time properties satisfied by a given model are satisfied in the same model under the control of any strategy, the critical section will always be protected. However, other fairness properties may depend on the scheduling policy. For instance, the property $\ctlAllw \ctlEvly \texttt{inCrit(1)}$ is not satisfied neither with the \texttt{blocked} policy nor with \texttt{roundRobin}.
\begin{maudexec}[mathescape]
Maude> red modelCheck(initial(4, p),
                      [] <> inCrit(1), 'blocked) .
rewrites: 7
result ModelCheckerResult: counterexample($\ldots$, $\ldots$)
\end{maudexec}
However, the counterexample consists only of 5 states instead of the 39 obtained with the standard model checker, and they are easier to understand in that they obey the restrictions of the strategy. They both show a process executing its loop continuously because it is never blocked. The preemptive version of round-robin makes the property hold.
\begin{lstlisting}[language={}]
$ umaudemc check semaphore.maude "initial(4, p)" \
           "[] <> inCrit(1)" "roundRobin(nil, 5, 5)"
The property is satisfied (1621 system states,
                           37735 rewrites).
\end{lstlisting}
We have used the \texttt{umaudemc} interface, since it allows calling \texttt{roundRobin} with arguments without declaring a new strategy. A time slice of $5$ has been fixed and the initial process list is empty. These parameters are immaterial to the satisfaction of the property since all processes are identical, but their values may affect the size of the model. We could have fixed the process order with the strategy \texttt{roundRobin(1 2 3 4, 5, 5)} instead, and the model would only have 90 states.

	Replacing the \texttt{p} program by \texttt{pIo}, which includes a blocking input/output operation, changes the situation. Thanks to the blocking operation, the \texttt{roundRobin} strategy is enough to ensure fairness since no process is left with the monopoly on the processor.
\begin{lstlisting}[language={}]
$ umaudemc check semaphore.maude "initial(4, pIo)" \
           "[] <> inCrit(1)" "roundRobin(nil)"
The property is satisfied (705 system states,
                           6199 rewrites).
\end{lstlisting}
However, the \texttt{blocked} policy may never activate a given process, because the next one that obtains the processor when the active process is blocked is chosen nondeterministically. We obtain a counterexample where the processes 2, 3 and 4 are being executed in turns repeatedly.
\begin{lstlisting}[language={},escapechar=^]
$ umaudemc check semaphore.maude "initial(4, pIo)" \
           "[] <> inCrit(1)" blocked
The property is not satisfied (18 system states,
                               73 rewrites).
^$[\ldots]$^
\end{lstlisting}

	This example could be expanded to support more realistic models and scheduling policies.

\subsection{The strategy language semantics as a strategy-controlled system} \label{sec:opsemimp}

	Strategies are useful to specify semantics of programming languages. A classical example is the $\lambda$-calculus and its different evaluation strategies to decide which redexes are reduced first, like call-by-value and call-by-name, which are also meaningful for similar functional languages, as we have considered in~\cite{pssm}. The Maude strategy language has also been used to specify and experiment with the semantics of a parallel extension of Haskell called Eden~\cite{eden}, and proposed as a general tool to define modular structural operational semantics~\cite{operational} that may easily include ordered rules or negative premises. For instance, these latter features appear in logic programming languages with negation and cut, like Prolog~\cite{maude30}.

	This example is a straightforward specification with strategies of the Maude strategy language small-step operational semantics presented in~\cref{sec:mcslang}, which can be used to model check any strategy-controlled system using the strategy-aware model checker with a fixed strategy. Strategies at the semantics level are in charge of handling the negative case of the conditional and specifying the relations $\to_c$, $\to_s$ and $\opsem$. Obviously, this approach is not recommended to model check strategy-controlled systems in practice, since checking them directly will be much more efficient, but we hope it will be useful to illustrate the usage of strategies to specify semantics without introducing a new language, and to clarify the semantics in~\cref{sec:mcslang} and its relation to model checking. Moreover, the example may be used to experiment with extensions of the strategy language or the model checker. 

	The syntax and semantics of the Maude strategy language depends essentially on the target system module being controlled. Hence, the specification of its small-step operational semantics should be parametric on it. Terms, strategies, modules, substitutions and so on are represented at the metalevel as declared in the predefined \texttt{META-LEVEL} module to simplify the specification and usage of the semantics. The sort \texttt{Term} of terms, \texttt{Strategy} of strategies, and \texttt{Module} of modules, as well as the different \emph{descent functions} that allow manipulating them efficiently like \texttt{metaApply} and \texttt{metaMatch}, are described in detail in the Maude manual~\cite{maude}. Thus, the parameter of the specification can be formalized in the following \texttt{MODULE} theory:
\begin{lstlisting}
fth MODULE is
	protecting META-MODULE .
	op M : -> Module .
endfth
\end{lstlisting}
As described in~\cref{sec:mcslang}, we must specify the \emph{execution state} terms, the rules in~\cref{sec:opsem}, and some strategies.
Execution states are described as terms of sort \texttt{ExState} using auxiliary sorts like \texttt{CtxStack} for stacks of pending strategies and variable contexts, with the empty-stack symbol \texttt{eps} ($\varepsilon$); and \texttt{SubtermSoup} for the substates of the subterm states.
\begin{lstlisting}
sorts ExState ExStatePart SubtermSoup
      SolutionSoup CtxStack .
subsort Term < ExStatePart .
subsort SolutionSoup < SubtermSoup .

op _@_ : ExStatePart CtxStack -> ExState [ctor] .
op subterm : SubtermSoup Term -> ExStatePart [ctor] .
op rewc : Term ExState Substitution Condition
          StrategyList CtxStack Term
          Context Term -> ExStatePart [ctor frozen] .

subsort Strategy < CtxStack .
op ctx : Substitution -> CtxStack [ctor] .
op eps : -> CtxStack [ctor] .
op __  : CtxStack CtxStack 
           -> CtxStack [ctor assoc id: eps] .

op _:_ : Variable ExState -> SubtermSoup [ctor] .
op _,_ : SubtermSoup SubtermSoup 
           -> SubtermSoup [ctor assoc] .

mb (V : T @ eps) : SolutionSoup .
op _,_ : SolutionSoup SolutionSoup
           -> SolutionSoup [ctor ditto] .
\end{lstlisting}
The subtype \texttt{SolutionSoup} of \texttt{SubtermSoup} contains those soups in which all nested states are solutions \texttt{T @ eps}. The term projection $\cterm : \xs \to T_{\Sigma}$ is described equationally:
\begin{lstlisting}
vars T P R RR : Term .         var V : Variable .
var  Ctx : Context .           var C : CtxStack .
vars Sb Th : Substitution .    var SL : StrategyList .
var  Sbs : SubtermSoup .       var SlS : SolutionSoup .
vars A B G : Strategy .        var C : EqCondition .
var  UPS : UsingPairSet .      var CS : CallStrategy .
var  X  : ExState .            var M : Module .

op cterm : ExState -> Term .
eq cterm(T @ C) = T .
eq cterm(rewc(V, X, Sb, C, SL, Th, R, Ctx, T) @ C) =  T .
eq cterm(subterm(SbS, T) @ C) = applySubs(T, ctermSubs(SbS)) .

op ctermSubs : SubtermSoup -> Substitution .
eq ctermSubs(V : X) = V <- cterm(X) .
eq ctermSubs((V : X), SbS) =
     ctermSubs(V : X) ; ctermSubs(SbS) .
\end{lstlisting}
where the \texttt{applySubs} function applies a substitution to a term, and \texttt{ctermSubs} builds the substitution from the variables of the \skywd{matchrew} to the current subterm being rewritten.

	The semantic rules in~\cref{sec:opsem} are represented almost directly as Maude rules. Their complete relation can be found in the source file~\cite{stratweb}, so here we will only show some of them. Notice that control rules are labeled with \texttt{ctl} and system rules with \texttt{sys} so that strategies can distinguish them later.
\begin{lstlisting}[escapechar=^]
rl  [ctl] : T @ ^idle^ S => T @ S .
rl  [ctl] : T @ (A *) S => T @ S .
rl  [ctl] : T @ (A *) S => T @ A (A *) S .
crl [ctl] : T @ (^match^ P ^s.t.^ C) S => T @ S
 if metaMatch(M, P, T, C, 0) :: Substitution .
\end{lstlisting}
Other rules are defined using auxiliary operators, either predefined like \texttt{metaMatch} in the \texttt{match} operator rule, or written for the occasion like in the following rules:
\begin{lstlisting}[escapechar=^]
crl [ctl] : T @ (^matchrew^ P ^s.t.^ C ^by^ UPS) S
         => subterm(subtermSoup(UPS, Sb),
              putInContext(applySubs(P,
                 removeVarsFromSb(Sb, UPS)),
              Ctx)) @ S
if {Sb, Ctx} |> MPS := metaMatch(M, P, T, C) .
rl [ctl] : subterm(SlS, T) @ S =>
           applySubs(T, ctermSubs(SlS)) @ S .
\end{lstlisting}
The first rule initiates a \texttt{subterm} state for the \skywd{matchrew} and builds all its components, and the second one concludes the subterm rewrite execution when all their substates are solutions, since \texttt{SlS} is a variable of sort \texttt{SolutionSoup}. Since the semantics of rewriting logic itself allows rules to be applied inside subterms, the rules that apply steps inside substates in~\cref{sec:opsem} are not needed. The same would be applied to the \texttt{rewc} operator for rule rewriting conditions, but both control and system transitions inside its substate should be considered control transitions for the whole \texttt{rewc} state, so the \texttt{frozen} attribute is added to the operator declaration --which prevents implicit rewriting inside its arguments-- and rules are applied explicitly using the following \texttt{rewc} rule with its rewriting condition controlled by a strategy, as we will see soon.
\begin{lstlisting}
crl [rewc] : rewc(P, X, Sb, C, SL, CS, RR, Ctx, ST) =>
    rewc(P, Y, Sb, C, SL, CS, RR, Ctx, ST) if X => Y .
\end{lstlisting}
Another interesting rule is that of strategy calls, which uses the auxiliary function \texttt{metaStratDefs} to calculate the matching contexts of the instantiated call term into the definitions of the module. These are returned as a \texttt{|>}-separated set, so that the rule selects one of them nondeterministically.
\begin{lstlisting}
crl [ctl] : T @ Q[[TL]] S => T @ CS S if CS |> CSS :=
 metaStratDefs(M, Q[[reduced(applySubs(TL, vsubs(S)))]]) .
eq ctx(Sb) ctx(Th) = ctx(Sb) .
\end{lstlisting}
The previous equation implements the tail-recursive call optimization, by removing the lowest of any pair of consecutive contexts in the stack.
Rule applications are handled using an overloaded \texttt{metaXapply} function that collects as a set the results of the builtin \texttt{metaXapply} descent function. The values in the initial substitution \texttt{Sb} are instantiated with the variables of the context and reduced.
\begin{lstlisting}
crl [sys] : T @ Q[Sb]{empty} S => T' @ S if T' |> TS :=
  metaXapply(M, T, Q, reduced(applySubs(Sb, vsubs(S)))) .
\end{lstlisting}
When strategies for rewriting conditions are specified, the state is rewritten to a \texttt{rewc} execution state, but we refer the interested reader to the complete specification for the details.

	On top of all these rules, strategies are used to specify the $\to_s$, $\to_c$, $\to_{s,c}$, $\opsem$ relations, and the $\opsem^*$ search for solutions that have been extensively used in~\cref{sec:mcslang}. Their definitions are simple:
\begin{lstlisting}
strats ->s ->c ->sc ->> opsem @ ExState .

sd ->>  := ->c * ; ->s .
sd ->sc := ->s | ->c .
sd ->c  := ctl | else{not(->sc* ; match T @ eps)}
               | rewc{->sc} .
sd ->s  := sys .
\end{lstlisting}
The definition of the control transition \texttt{->c} includes two other labels in addition to \texttt{ctl}. One is the rule \texttt{rewc} that applies transitions inside the substate of a \texttt{rewc} state, which should be considered control steps no matter if they are in the substate, as explained before. For that reason, the strategy applied to the substate is $\to_{s,c}$. The other label, \texttt{else}, refers to the rule for the negative-branch rule of the conditional, defined as
\begin{lstlisting}
crl [else] : T @ (A ? B : G) S
          => T @ G S if T @ A vctx(S) => X [nonexec] .
\end{lstlisting}
Its rewriting condition is controlled by a strategy that fails if \texttt{->sc* ; match T @ eps} succeeds, in other words, if a solution is reachable from \texttt{T @ A vctx(S)}, as required by the original rule.
Finally, the \texttt{opsem} definition
\begin{lstlisting}
sd opsem := test(->c * ; match T @ eps)
              ? idle : ->> ; opsem .
\end{lstlisting}
captures the requirements of the strategy-controlled model described in~\cref{def:mslstrat}: it allows both infinite executions of \texttt{->>} transitions, and finite ones ending in states where a solution can be reached by control transitions. Ensuring that the \texttt{->>} transition is seen as the atomic step, for what the opaque strategy feature described in~\cref{sec:opaque} can be used, the system controlled by \texttt{opsem} from the initial state \texttt{$\overline t$ @ $\overline \alpha$} is equivalent to $t$ controlled by $\alpha$ modulo the \texttt{cterm} projection. The \skywd{matchrew} combinator is executed without bias, but a biased version can be programmed with strategies using \skywd{matchrew}, insisting in the reflective nature of this example.

	The last requirement for model checking is defining atomic propositions. Since states and strategies have been represented at the metalevel, atomic propositions are also represented as metaterms.
\begin{lstlisting}
mod NOP-PREDS{X :: MODULE} is
	protecting NOP-RULES{X} .
	including SATISFACTION .

	subsort ExState < State .
	op prop : Term -> Prop [ctor] .

	var XS : ExState .
	var P  : Term .
	eq XS |= prop(P) = getTerm(metaReduce(M,
	           '_|=_[cterm(XS), P])) == 'true.Bool .
endm
\end{lstlisting}
The predicate term is wrapped in a \texttt{prop} symbol, whose satisfaction is defined using the predefined \texttt{metaReduce} function that evaluates \texttt{cterm($q$) |= $p$} in the base module, where $q$ and $p$ are the terms metarepresented by \texttt{XS} and \texttt{P} respectively. The \texttt{NOP-PREDS} module is parameterized by the \texttt{MODULE} theory, which determines the underlying module.

	Finally, we can instantiate the semantics with the philosophers' example. The formal constant \texttt{M} in the \texttt{MODULE} theory is mapped to the metarepresentation of the \texttt{DINNER-MCS} module obtained with the builtin \texttt{upModule} operator.
\begin{lstlisting}
view Philosophers from MODULE to META-LEVEL is
	op M to term upModule('DINNER-MCS, true) .
endv
\end{lstlisting}
To model check the formula $\ctlAllw \ctlEvly \bigvee_{k=1}^4 \texttt{eats($k$)}$ from the initial term \texttt{initial} using the \texttt{parity} strategy, we only have to model check the execution state \lstinline[mathescape]{$\overline t$ @ $\overline \alpha$} combining the metarepresentations of \texttt{initial} and \texttt{parity} against the property with the atomic propositions replaced by their metarepresentations inside the \texttt{prop} symbol. The semantics is executed under the control of \texttt{opsem} with \texttt{->>} as opaque strategy to respect the transitions of the original model.
\begin{maudexec}
Maude> red modelCheck(reduced('initial.Table) @ 'parity[[empty]], 
 [] <> (prop('eats['0.Zero]) \/ ...
        \/ prop('eats['s_^4['0.Zero]])), 'opsem, '->>) .
rewrites: 497449
result Bool: true
\end{maudexec}

\section{Implementation} \label{sec:implementation}

	The strategy-aware model-checker implementation is based on the operational semantics of~\cref{sec:opsem} and it relies on two existing resources: the Maude LTL model checker and the C++ infrastructure for the execution of strategies.

	As stated in~\cref{sec:ltlmc}, the Maude model checker is an optimized implementation of the standard explicit-state LTL algorithm explained in~\cref{sec:ltl}, composed of three distinct parts: a generator of Büchi automata from LTL formulae, an on-the-fly generator of the automaton that represents the state and transition structure of the model, and the nested depth-first search algorithm that finds a counterexample on the intersection of the two automata. Since we maintain the property logic, and thanks to the low coupling of the three components in the original implementation, it has only been necessary to modify the second of them. The model is presented in C++ as a collection of states indexed by natural numbers whose successors can be queried and calculated on-the-fly using a \texttt{getNextState} method. Each state is associated to a term, in which atomic properties can be checked. The states of the original model consist merely of a term, but the strategy-aware model must incorporate the strategy execution state.

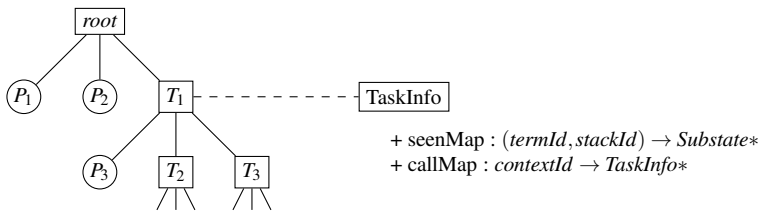
\begin{figure}[h]\centering
\begin{tikzpicture}[process/.style={draw, circle, inner sep=1pt}, task/.style={draw}]
	\node[task] (RT) at (0, 0) {$\mathit{root}$};
	\node[task] (TI) at (4, -1) {TaskInfo};
	\node[task] (T1) at (1, -1) {$T_1$};
	\node[task] (T2)  at (1, -2) {$T_2$};
	\node[task] (T3) at (2, -2) {$T_3$};
	\node[process] (P1) at (-1, -1) {$P_1$};
	\node[process] (P2) at (0, -1) {$P_2$};
	\node[process] (P3) at (0, -2) {$P_3$};

	\node (TIinfo) at (6.25, -1.75) {\small\begin{tabular}l
		+ seenMap : $(\mathit{termId}, \mathit{stackId}) \to \mathit{Substate*}$ \\
		+ callMap : $\mathit{contextId} \to \mathit{TaskInfo*}$
	\end{tabular}};

	\draw (RT) -- (T1);
	\draw (RT) -- (P1);
	\draw (RT) -- (P2);
	\draw (T1) -- (P3);
	\draw (T1) -- (T2);
	\draw (T1) -- (T3);

	\draw (T2) -- (0.75, -2.5);
	\draw (T2) -- (1, -2.5);
	\draw (T2) -- (1.25, -2.5);
	\draw (T3) -- (1.75, -2.5);
	\draw (T3) -- (2, -2.5);
	\draw (T3) -- (2.25, -2.5);

	\draw[dashed] (T1) -- (TI);
\end{tikzpicture}

\caption{Example hierarchy of tasks (boxes) and processes (circles) with a task info.} \label{fig:tasks}
\end{figure}

	The calculation of the successors of a state uses the strategy execution infrastructure of the \texttt{srewrite} and \texttt{dsrewrite} commands, in whose implementation we have contributed. This is supported in a collection of tasks and processes, which have been slightly and conveniently adapted. Different classes of processes are in charge of applying rules, finding pattern matches and testing conditions, executing strategy definitions, decomposing strategies and processing their arguments\ldots{} for what they may create and destroy new processes and tasks. These processes are kept in a global double-linked list and executed in a round-robin or FIFO policy by the \texttt{srewrite} or \texttt{dsrewrite} command respectively. Each process is also attached to a task, and in turn, these are organized hierarchically as a tree (see~\cref{fig:tasks}). Tasks group processes being responsible for the same subsearch (which may appear in the evaluation of rule rewriting conditions, of the condition of conditional operators\ldots) and also delimit variable environments produced by the \texttt{matchrew} operator or strategy calls. Moreover, each task maintains a set of visited term-strategy pairs to avoid repeating unnecessary calculations and to let the search terminate in the presence of cyclic executions. The visited set of each task is independent, because the same strategy could be applied to the same term but with other values for the variables or in a different subsearch. The pending strategies are handled by a queue similar to those of the operational semantics, and in fact the strategies of the term-strategy pairs are indices to this structure. Each task additionally holds the index of the pending strategies to be executed for each solution of the subsearch it hosts.

	For the model checker, light but essential changes are applied to this structure. First, the global list of processes is split into multiple lists local to each model state to allow calculating and identifying the successors of any chosen state. Each model state stores a pointer to the current process in its list, which is executed in round-robin. According to the $\opsem$ semantics, a new state is only generated when a rewrite takes place (or an opaque strategy yields a result, see~\cref{sec:opaque}) and the active process notifies it to the object in charge of managing the model graph. At this moment, checking whether the new state has been visited before is crucial to ensure the termination of the algorithm in the conditions indicated in~\cref{sec:maudesmc}, and doing it safely and efficiently is perhaps the most complicated aspect of the implementation. Ideally, two model states are equivalent if they correspond to the same execution state of the operational semantics. Checking the equivalence just at the state creation is enough not to lose any cycle, but actually, a model-checker state visits many states of the semantics, always related by control transitions, when executing its list of processes.
Some of them may be as general as the formal state represented by the initial process, having the same successors by the $\opsem$ transition, but others may have lost continuations because of a rule like $\alpha \seq \beta \to \alpha$. In order to anticipate the detection of cycles, with the consequent advantages in execution time and simplicity of the possible counterexamples, the model controller executes as many \emph{conservative} operations as possible to compare with a simpler instance of the state, it generates \emph{substates}\footnote{Substates are entirely similar to states except that they are not part of the model, and consequently they are not linked as successors by other (sub)states, but as \emph{dependencies}, from which successors are copied instead.} to reuse the search from non-conservative branches of the states when convenient too, and merges states if their equivalence is detected afterwards. The correspondence from an implementation state to a state of the semantics is based on adding to the subject term $t$ being rewritten by the current process the pending strategies according to the strategy stack index $t \ao \vec\alpha$, the variable environment and continuation of the enclosing parent task $t \ao \vec\alpha \theta \vec\beta$, and the appropriate execution state constructor like $\subterm(\ldots, x_i : t \ao \vec\alpha, \ldots)$ according to the parent task too. Hence, checking if two model states are equivalent goes through comparing their subject terms, their pending strategy indices, and their ancestor tasks. The first two were already compared in the normal execution using the task-local visited sets, but this is insufficient for several reasons. On the one hand, aborting the search when detecting a visited state is not an option here because we must know how the execution continued to complete the graph, so the visited set should be replaced by a table. On the other hand, as per the tail-recursive call optimization described in~\cref{sec:opsem}, the model can be finite even in the presence of nonterminating strategy calls if these are tail recursive with finitely many different arguments. The execution infrastructure does not compare the arguments of the strategy calls and generates a different task for each call, and so this circumstance is not detected.  Both problems are solved associating to each task a \emph{task info} structure (see~\cref{fig:tasks}) holding the aforementioned table, which maps each term-strategy pair to the substate that continues its execution, and another table associating variable environments to the \emph{task info} structure shared by all the recursive strategy call tasks starting there. Except for this case, the state comparison is done locally at the task level and this may delay the detection of cycles in some cases. For example, if $\beta$ is \texttt{\skywd{matchrew} $x$ \skywd{by} $x$ \skywd{using} $\texttt r$} and $\texttt r$ is a rule that rewrites \texttt{a} to \texttt{b} and \texttt{b} to \texttt{a}, a cycle like
\begin{align*}
\subterm(x : \texttt a \ao \texttt r; x) \ao \beta^* &\opsem \subterm(x : \texttt b \ao \varepsilon, x) \ao \beta^* \to^*_c \texttt b \ao \beta^* \\
	& \opsem \subterm(x : \texttt a \ao \texttt r; x) \ao \beta^*
\end{align*}
will not be detected in its final state. The reason is that the tasks for the first and last $\subterm$ are not the same: the first has been destroyed when the execution of the \texttt{matchrew} has finished and the second is a new one with a fresh table of visited pairs. However, no cycle will be missed in this situation or a similar one, because the execution must evolve to a lower level in the task hierarchy, in this case to $\texttt b \ao \beta^*$, when the parent task will be the same and the cycle will be detected. Not to miss any such case, the visited table is always looked up when an execution descends to a parent task. Obviously, a deeper comparison of the tasks could prevent this inconvenience at a higher cost. A compromise should be found between state-space reduction, speed and memory required for each state, always ensuring that the algorithm finishes when the abstract execution states are finite.

	All things considered, the model-checker states represented in C++ correspond to states of the operational semantics, and in particular with those reachable from the initial state $t \ao \alpha$ by the ${\opsem} = {\to_c^*} \circ {\to_s}$ transition, which connects all of them except when it comes to opaque strategies. The cycle detection mechanism ensures that the algorithm terminates under the assumptions of~\cref{sec:maudesmc}.

\section{Evaluation} \label{sec:evaluation}

The model checker presented in this article has been tested with several examples of temporal properties on strategy-controlled specifications available in the Maude strategy language web page~\cite{stratweb}. Since ours is the only model checker available for strategy-controlled systems, we cannot easily compare the performance of the tool with other implementations, except by translating the models to the potentially very different formalisms used by them. However, our tool is an extension of the Maude LTL model checker, with which it shares all of its components except those related with the system automaton. Hence, measuring the executions of both model checkers for pairs of strategy-controlled and equivalent rule-only Maude specifications would be a targeted and significant comparison. From the user point of view, the question is whether applying our model checker on a strategy-controlled specification is more convenient than translating that specification to be model checked by the standard tool.
We think that the results in this section answer positively to this question, since the performance penalties that may appear in some cases are not significant enough to renounce to the advantages of using strategies or to spent time translating the specification to the rule-only subset of Maude. Moreover, we think that the availability of this model checker makes the introduction of strategies more convenient in suitable specifications aimed to be verified, which was previously discouraged by the absence of such a tool.
Using strategies to specify systems does not pursue a performance improvement, but obtaining clearer specifications and experimenting more easily with them, so this additional abstraction may occasionally have some manageable cost, which we keep trying to reduce as much as possible. However, strategies may help to improve the efficiency of specifications without complicating them, as we have seen in~\cite{memstratmc}.

	We have translated the strategy-based specification of the examples in~\cref{sec:philosophers,sec:scheduling} to distinct rule-base ones for each strategy. In the first case, we have also specified the example in the Promela language and model checked it using the well-known Spin model checker~\cite{spinmc}.
The changes in the data representation and the rules that have replaced the strategies are as or ever more efficient than the original strategies, but the specifications are more obscure and need to be different for each control mechanism. Similarly, by translating the first example specification to multiple lower-level models for the Spin model checker, the performance has improved at some cost in readability. Writing implementations for every strategy of the second model in Spin would be a much harder work. Moreover, we have lost an interesting property that strategies provide for free, that the behaviors of the controlled model are a subset of those of the uncontrolled one.

	Another comparison between different model checkers operating on the low-level Kripke structure produced by the extension in this paper is available in~\cite{btimemc}. In both cases, we have evaluated strategies as a specification resource, but they can also be used for the only purpose of improving the performance of the verification, by restricting the execution space or conducting the model checker towards conjectured counterexamples. This interesting study is left for future work.

\subsection{The philosophers problem} \label{sec:philperf}

	As we mentioned in~\cref{sec:philosophers}, the dining philosophers problem can be generalized to $n$ philosophers and $n$ forks without modifying its terms, rules, and strategies. Only the initial term and the temporal formulae have to be adapted, but they have been defined so that the number of philosophers is received as a parameter. \Cref{table:phil} shows under the SL columns the number of states, the time in milliseconds, the number of rewrites, and the peak usage of heap memory spent to model check the two considered LTL properties in the strategy-controlled specification with an increasing number of philosophers. As a reference, the number of states in the uncontrolled system is $3^n$.
The last row for the \texttt{parity} strategy is empty since the model checker does not finish in reasonable time for that number of states. All measures grow exponentially as the number of states, including the amount of memory used for the first property, which reaches $2.12$ Gb for $n=21$ and becomes unfeasible for $n=23$. On the contrary, the memory peak using the \texttt{turns} strategy stays low and stable.

\begin{table}\centering
\leavevmode\kern-2em\begin{tabular}{c  r  r r r  r r  r r r}
	\toprule
	Num	& \multicolumn1c{States} & \multicolumn3c{Time (ms)} & \multicolumn2c{Rewrites} & \multicolumn3c{Memory peak (Mb)} \\
	\cmidrule(lr){2-2} \cmidrule(lr){3-5} \cmidrule(lr){6-7} \cmidrule(lr){8-10}
	phil 	& \multicolumn1c{All}
				& SL	& Maude & Spin	& SL 		& Maude		& SL 		& Maude 	& Spin \\
	\midrule
	3	& 12		& 37	& 37	& 1161	& 118		& 99		& 8.17		& 8.14		& 128.8 \\
	5	& 48		& 38	& 37	& 1187	& 548		& 493		& 8.21		& 8.16		& 128.8 \\
	7	& 180		& 43	& 39	& 1235	& 2354		& 2191		& 8.58		& 8.48		& 128.8 \\
	11	& 2268  	& 160	& 68	& 1286	& 36962		& 35503		& 11.76		& 10.39		& 128.8 \\
	13	& 7776		& 537	& 168	& 1329	& 139316 	& 134941	& 20.65		& 14.44		& 128.8 \\
	17	& 87480		& 8876	& 2216	& 1357	& 1.87e6 	& 1.83e6	& 178.25	& 107.22	& 128.8 \\
23	& 3.07e6	& -	& -	& 4029	& -		& -		& -		& -		& 550.4			\\
	27	& 3.19e7	& -	& -	& 40299	& -		& -		& -		& -		& 5241.8			\\
	\bottomrule
\end{tabular}

	\medskip
	(a) Someone eats $\ctlEvly \bigvee_{k=0}^{n-1} \texttt{eats}(k)$ with \texttt{parity}
	\bigskip

\begin{tabular}{c  r r r  r r r  r r}
	\toprule
	Num	& \multicolumn3c{States} & \multicolumn3c{Time (ms)} & \multicolumn2c{Rewrites} \\
	\cmidrule(lr){2-4} \cmidrule(lr){5-7} \cmidrule(lr){8-9}
	phil 	& SL	& M	& Spin	& SL	& Maude	& Spin	& SL	& Maude 	\\
	\midrule
	3	& 10	& 9	& 66	& 38	& 37	& 1160	& 137	& 137 		\\
	5	& 16	& 15	& 170	& 38	& 38	& 1187	& 541	& 553		\\
	7	& 22	& 21	& 332	& 41	& 40	& 1235	& 2077	& 2109		\\
	11	& 34	& 33	& 770	& 89	& 89	& 3987	& 31981	& 32077		\\
	13	& 40	& 39	& 1066	& 245	& 245	& 27596	& 1.27e5 & 1.27e5 	\\
	17	& 52	& 51	& -	& 3387	& 3418	& -	& 2.03e6 & 2.03e6	\\
	23	& 70	& 69	& -	& 234431& 229617 & -	& 1.3e8 & 1.3e8		\\
	\bottomrule
\end{tabular}

	\medskip
	(b) All eat $\ctlAllw \bigwedge_{k=0}^{n-1} \ctlEvly \texttt{eats}(k)$ with \texttt{turns}
	\bigskip

\caption{Execution measures for the philosophers problem using Maude and Spin.} \label{table:phil}
\end{table} 

	We may inquire whether a better performance could be obtained if instead of specifying these restrictions as strategies we modify the system module so that rules incorporate them, albeit the other advantages of strategies would be lost. In the case of the \texttt{parity} strategy, the \texttt{left} and \texttt{right} rules are implemented by the following five rules:

\begin{lstlisting}[literate={psi}{\psiCompat}1{=>}{{=>}}2, escapechar=^]
crl [left-even]  : psi (o | Id | o)
                => (psi | Id | o) if 2 divides Id .
rl  [left-odd]   : psi (o | Id | psi) => (psi | Id | psi) .
crl [right-odd]  : (o | Id | o) psi
                => (o | Id | psi) if ^not^(2 divides Id) .
rl  [right-even] : (psi | Id | o) psi => (psi | Id | psi) .
rl  [left-even]  : < (o | Id | o) L psi >
                => < (psi | Id | o) L > .
\end{lstlisting}
The \texttt{turns} strategy has also been implemented without strategies by using a token passed to the next philosopher within the rules. Under the Maude columns of~\cref{table:phil}, there are the results of checking the same properties using the standard model checker on the transformed specifications. In the \texttt{parity} case, the number of states does not change and the other measures are lower in the transformed system. However, the critical number in which verification is not longer feasible coincides (the modified system takes $1.20$ Gb with $n=21$). In the case of \texttt{turns}, the figures are equivalent or even better for the original specification. No more than 8.7 Mb of memory are used both with and without strategies.
Hence, at least for this problem, there is no significant performance loss on using strategies. The greater usage of memory of the strategy-aware model checker can be explained by a second cache of the evaluation of atomic propositions on states in addition to that already provided by the common infrastructure. In general, although not in this case, different states of the strategy-controlled model may represent the same term, and this cache tries to avoid the evaluation of the same property not only on the same state, but on the same term. This feature can be disabled at compile time to reduce the memory consumption.

	We have also specified this same problem in the Promela language of the Spin model checker~\cite{spinmc}. The model consists of two byte arrays of length $n$ describing the availability of each fork and the number of forks retained by each philosopher, which are updated by a process for each philosopher in a loop that implements the \texttt{parity} restriction or the \texttt{turns} strategy using an auxiliary variable for the current turn. The verification process in Spin consists of generating a C verifier from the Promela specification and the LTL formula using the \texttt{spin -a} command, compiling it with the C compiler, where we have used the \texttt{-O2} optimization flag, and running the resulting program. The measures of the execution of the last binary are included in~\cref{table:phil}, showing that its performance is noticeably better in the \texttt{parity} case. While both Maude specifications cannot handle in reasonable time and memory limits the size $n=23$, Spin verifies this case in two seconds and can reach up to 27 philosophers with 12 Gb of RAM.\footnote{The fixed value of 128.8 for the memory usage of Spin in the smaller cases is due to a hash table reserved by the model checker in its default setting, which we have not changed.}
On the contrary, its behavior for the \texttt{turns} strategy is much worse. Once generated, the execution time of the verifier is small, but the first phase's time quickly grows due to the processing of the temporal formula. We have interrupted the \texttt{spin -a} command for $n=17$ after ten minutes, while this case can be checked in less that 4 seconds in Maude.

\subsection{Scheduling policies}

	The \texttt{roundRobin} strategy and its preemptive version in the example on scheduling policies in~\cref{sec:scheduling} have also been translated to rule-only Maude specifications, by extending the machine state.
\begin{lstlisting}[moredelim={[is][\itshape]{\#}{\#}}]
op {_,_,_,_} : Soup Memory #List{Pid} Mark# 
                 -> MachineState [ctor] .
op {_,_,_,_,_} : Soup Memory #List{Pid} Nat Mark#
                   -> MachineState [ctor] .
\end{lstlisting}
The list of process identifiers and the preemption counter maintained in the strategy arguments are stored in the machine state, which also includes a mark that will help to define the modified rules. Strategies allow using the failure of the execution of a process to switch to the next of the list, and this cannot be easily handled within the rules. In summary, we have solved the problem by modifying the rules where a process can get blocked to explicitly treat the negative case, switching to another process that can take a step.
\begin{lstlisting}
crl [exec] : {[I, wait(Q) ; R] | [J, P] | S, M, I J PL, G} 
          => {[I, wait(Q) ; R] | [J, P'] | S', M', PL', m(G)}
 if [Q, N] RM := M
 /\ N <= 0
 /\ pidsIn(S) subset list2set(PL)
 /\ {[J, P] | S, M, J PL I, pending}
 => {[J, P'] | S', M', PL', done} .
\end{lstlisting}
One of the conditions is that all processes in the soup are in the list of processes, since otherwise the \texttt{roundRobin} strategy would try giving the processor to missing processes first, for what another rule is required. The mark at the end of the state is to ensure that one and only one step (the first one of the new active processes) is executed in the rewriting condition.We hope that the reader will notice how much complex and error prone these terms and rules are compared to the strategies in the original specification. Moreover, the combination of the original \emph{exec} rules and the strategies are much more readable and understandable, since rules do not have to handle the next step of the processor. Finally, the same set of rules were valid for all control mechanisms, which are guaranteed to be a restriction of the uncontrolled model behavior, while multiple set of rules have been written for each policy without that guarantee.

	The results of the verification of the property $\ctlAllw \ctlEvly \mathit{inCrit}\,(1)$ for the initial states \texttt{initial($n$, $p$)} for $p \in \{\texttt{pIo}, \texttt{p}\}$ using the \texttt{roundRobin} and its preemptive version respectively are shown in~\cref{table:sched}. Note that these strategies fall in the worst cases of the implementation described in~\cref{sec:implementation}, where several \skywd{matchrew}s defer the detection of cycles, increasing the number of the model states. All measures decrease noticeably in the translated specification, but both become unmanageable for almost the same sizes.

\begin{table}\centering
\leavevmode\kern-2em\begin{tabular}{c c  r r  r r  r r  r r}
	\toprule
	$p$ & $n$ & \multicolumn2c{States} & \multicolumn2c{Time (ms)} & \multicolumn2c{Rewrites} & \multicolumn2c{Memory peak (Mb)} \\
	\cmidrule(lr){3-4} \cmidrule(lr){5-6} \cmidrule(lr){7-8} \cmidrule(lr){9-10}
			&	& SL	 	& Maude		& SL	 	& Maude		& SL 		& Maude		& SL 		& Maude 	\\
	\midrule
	\texttt{pIo}	& 4	& 705		& 321 		& 73		& 68 		& 6201		& 3672		& 10.12		& 9.88		\\
			& 6	& 28501		& 9781 		& 505		& 214 		& 457301	& 191176	& 20.6		& 15.34		\\
			& 9	& 1.98e7	& 4.93e6	& 8.52e5	& 1.91e5	& 6.24e8	& 1.79e8	& 8765.37	& 3092.26	\\
	\texttt{p}	& 4	& 1621		& 825		& 99		& 84 		& 37737		& 17568		& 10.73		& 10.17		\\
			& 6	& 71107		& 24901		& 3163		& 762 		& 4.47e6	& 1.10e6	& 52.8		& 23.37		\\
			& 8	& -		& 1.39e6	& -		& 90263		& -		& 1.11e8	& -		& 828.56		\\
	\bottomrule
\end{tabular}
\caption{Execution measures for the scheduling policies example.} \label{table:sched}
\end{table}

	Although model checking the strategy-controlled system provides a worse performance in this case, strategies are still useful for their greater flexibility. However, as future work, we should consider updating the implementation to improve how \skywd{matchrew}s are handled and its performance.

\section{Conclusions and future work}

	Strategies are a useful resource to build compositional rewriting-based specifications, where the control of rule application is described separately without obscuring the data representation of the model and the rules themselves. In order to make the use of strategies worth, models thus described should count with similar verification facilities as their uncontrolled counterparts, being model checking one of the most spread techniques. Making model checking meaningful for strategy-controlled systems is based on a clear and simple principle: strategies limit the possible executions of a model, and so properties should only refer to the allowed behaviors. A general procedure to model check these systems is transforming them to plain Kripke structures where standard algorithms can be applied.

	The newest versions of the Maude specification language already come with an LTL model checker and an object-level strategy language to control rewriting. Using the strategy language implementation, we have extended the builtin LTL model checker to support strategy-controlled systems. In order to formalize which executions are allowed by a strategy expression, a small-step operational semantics of the language is defined, which can be used to construct the transformed strategy-aware Kripke structure that inspires the actual implementation. While the Maude strategy language is Turing-complete, model checking is only decidable if this transformed structure is finite, or equivalently, if the set of allowed traces is an $\omega$-regular language. Sometimes decidability can be concluded from syntactical features of the strategy expressions and other considerations.
This paper describes three examples of strategy-based specifications related to concurrency problems and language semantics where temporal properties have been checked, and others are available in the strategy language web page and other papers~\cite{stratweb,metatrans,memstratmc,bitmlmc}. The performance of the model checker has been compared using these examples, and from our point of view, the benefits of the high-level specification resource of strategies deserve the occasional additional cost caused by them.

	In other work~\cite{btimemc}, we have extended the strategy-aware model checker to support branching-time logics like CTL* and $\mu$-calculus. These are checked using external tools through the \textsf{umaudemc} program, which facilitates checking LTL properties too. This work can be extended in several other directions like the verification of non-closed or non-intensional strategies that are able to capture fairness constraints and require other approaches, and the study of the associated satisfaction problem in relation with strategy or controller synthesis~\cite{controllerSynthesis} and the currently active research on strategic logics~\cite{mogaveroJournal}. Other formalisms could also be targeted like probabilistic and narrowing-based models. 
Apart from checking strategy-controlled specifications, the model checker can also be used to analyze classical models more efficiently with strategies that limit the state space or guide the search to the counterexamples that refute a property, which may probably be simpler and shorter. This is another worthy application of this model checker and it is currently being explored.
 
\begin{acknowledgements}
	This work was partially supported by the Spanish Ministry of Science and Innovation (PID2019-108528RB-C22). Rubén Rubio is partially supported by the Spanish Ministry of Universities (FPU17/02319).
\end{acknowledgements}

\section{Declarations}

\paragraph{Funding} This work was partially supported by the Spanish Ministry of Science and Innovation (PID2019-108528RB-C22). Rubén Rubio is partially supported by the Spanish Ministry of Universities (FPU17/02319).

\paragraph{Conflicts of interest}

	The authors declare that there is no conflict of interest.

\paragraph{Availability of data and material}

	The model checker, the examples introduced in this paper, and the material for the tests in~\cref{sec:evaluation} are available in~\url{https://maude.ucm.es/strategies}.

\paragraph{Code availability}

	The source code of the model checker is available at \url{https://github.com/fadoss/maudesmc}, and the source code of the examples is available in \url{https://maude.ucm.es/strategies}.

\bibliographystyle{plain}

\newpage\appendix
\section{Proofs}

\setcounter{lemma}{0}
\setcounter{proposition}{0}

\begin{proposition}
	Given $E \subseteq S^\omega$, there is a finite Kripke structure $\mathcal K'$ such that $\ell(\Gamma^\omega_{\mathcal K'}) = \ell(E)$ iff $\ell(E)$ is closed and $\omega$-regular.
\end{proposition}

\begin{proof}
	Notice that the finite Kripke structure $\mathcal K'$ can act as Büchi automaton and vice versa. Given a Kripke structure $(S, \to, I, AP, \ell)$, the automaton $(S \cup \{\iota\}, \mathcal P(AP), \delta, \iota, S \cup \{\iota\})$ with $\delta(\iota, P) = \{ s \in I : \ell(s) = P \}$ and $\delta(s, P) = \{ s' \in S : \ell(s') = P \;\wedge\; s \to s' \}$ is considered; and given an automaton $(Q, \mathcal P(AP), \delta, q_0, F)$, we consider the Kripke structure $(Q \times \mathcal P(AP), \to, \{ (s, \ell(s)) : s_0 \to s, s_0 \in I \}, AP, \pi_2)$ with $(s, P) \to (s', P')$ if $s \to s'$ and $\ell(s') = P'$. Checking that their word and execution coincides is straightforward, taking into account that $E$ is closed and so $F$ is irrelevant.
\end{proof}

\begin{proposition}
	The projection of the infinite traces of $\ltmsl$ by $\pi_1 \circ \cterm$ coincides with the stuttering-extension of $E(\alpha, t)$.
\end{proposition}

\begin{proof}
	Remember that $\ltmsl$ is defined as $(\xs \times \{0\} \cup \Sol \times \{1\}, \opsem_\Sol)$ where $(q, 0) \opsem_\Sol (q', 0)$ if $q \opsem q'$ and $(q, k) \opsem_\Sol (q, 1)$ if $q \in \Sol$ for $k=0,1$. Consequently, all infinite traces of $\ltssl$ are infinite traces of $\ltmsl$ (with a zero in the second component), but these are exactly $\mathrm{Ex}^\omega(\alpha, t)$ by definition. The finite traces $q_1 \cdots q_n$ in $\ltssl$ are finite traces $(q_1, 0) \cdots (q_n, 0)$ in $\ltmsl$, but if and only if $q_n \in \Sol$ they can be extended to the infinite traces $(q_1, 0) \cdots (q_n, 0) (q_n, 1) \cdots$. These are the traces in $\mathrm{Ex}^*(\alpha, t)$, whose stuttering-extended projections in $E(\alpha, t)$ are precisely $\cterm(q_1) \cdots \cterm(q_n) \, \cterm(q_n) \cdots$, the projection of the extended executions ending in a halting state. Thus, $\cterm(\pi_1(\Gamma^\omega_{\ltmsl}))$ is the stuttering-extension of $E(\alpha, t)$, where $\pi_1(x, y) = x$.
\end{proof}

\begin{lemma} \label{lem:decidable}
	If the underlying equational theory is decidable and the reachable states from $t \ao \alpha$ are finitely many, $\opsem$ and $\to_{s,c}$ are decidable.
\end{lemma}

\begin{proof}
	All the rules defining $\to_c$ and $\to_s$ but [else] are decidable, since they only involve immediate term manipulations, matching, substitution application, etc. The [else] rule is decidable on an execution state $t \ao \ifthel\beta\gamma\zeta s$ if the reachable states from $t \ao \beta \, \vctx(s)$ are finitely many. However, these are already embedded in the states reachable from the conditional, since $t \ao \ifthel\beta\gamma\zeta s \to_c t \ao \beta\gamma s$ and all the successors of $t \ao \beta \, \vctx(s)$ are successors of $t \ao \beta\gamma s$ with $\vctx(s)$ replaced by $\gamma s$, since the same rules can be applied with their free variables $s$ changed like this. In case of nested conditionals, this argument can be repeated from inside out conditional expressions, so all these states are reachable from the initial $t \ao \alpha$, and so they are finitely many and the rule is decidable. Since the reachable states are a finite set, deciding $q \opsem q'$ is finding a path via $\to_c$ transitions from $q$ to any predecessor of $q'$ by a $\to_s$ transition, so it is decidable.
\end{proof}

\begin{proposition}
	For any $\infty$-recursively enumerable language $L \subseteq \Gamma_{\mathcal M}$, there is some strategy expression $\alpha$ such that $E(\alpha) = L$.
\end{proposition}

\begin{proof}

	The finite-word part $L_*$ and the infinite-word part $L_\omega$ of $L$ can be considered separately. In effect, if there is a strategy expression $\alpha$ such that $E(\alpha) = L_*$ and a strategy expression $\beta$ such that $E(\beta) = L_\omega$, then $E(\alpha \disj \beta) = E(\alpha) \cup E(\beta) = L_* \cup L_\omega = L$.

	Let us start with the finite-word part $L_*$. Since it is recursively enumerable, there must be a Turing machine $M = (Q, \Gamma, T_\Sigma, q_0, F, \delta)$ such that $L_* = L(M)$. Turing machines can easily be represented in Maude, but for the sake of brevity we will see them as terms with two defined operators: \texttt{accept} that evaluates to \texttt{true} if the word in its tape is accepted, and \texttt{append} that puts a symbol on its tape. A generic specification including these functions is available at~\cite{stratweb}. The strategy that admits exactly $L_*$ is defined as a recursive expression that carries a Turing machine as an argument and fills the tape with the visited terms while rewriting. At some point, it runs the Turing machine to decide if the accumulated word is accepted and can be yielded as a solution of the strategy.
\begin{lstlisting}
sd climb(M, N) := run(M, N) | climb(M, s(N)) .
sd run(M, 0) := match S s.t. accept(append(M, S)) .
sd run(M, s(N)) := matchrew S by S using (all ; run(append(M, S), N)) .
\end{lstlisting}
The initial strategy call is \texttt{climb(M0, 0)} where \texttt{M0} is in its initial state with an empty tape. Observe that this nonterminating strategy \texttt{climb} fixes in advance the length of the executions to be recognized by \texttt{run}. This is a technical detail to avoid admitting infinite executions that are accumulation points of the finite words in the language. We claim that \texttt{run(M0, $n$)} admits all words in $L_*$ of length $n$. In effect, the contents of the tape of \texttt{M} is the sequence of terms visited until but not including the current subject term \texttt{S}. If the second argument is positive, the current state is appended to the tape by \texttt{append(M, S)}, a new rewrite step is performed, hence maintaining the invariant in the previous phrase, and \texttt{run} is called with $n-1$. If the counter is zero, the Turing machine \texttt{M} is executed by \texttt{accept(M')} after appending the last state \texttt{S}, which only evaluates to \texttt{true} if the word or execution in its tape is in $L_*$, and only in this case the strategy yields a solution. Finally, the strategy \texttt{climb} clearly admits the union of all executions allowed by \texttt{run(M, $n$)} for all $n \in \N$, which are all the bounded subsets of $L_*$, so it admits $L_*$. Moreover, it does not admit any other word since the infinite $\to_{s,c}$-execution repeating
\[ t \ao \texttt{climb(M, $n$)} \to_c t \ao (\texttt{run(M, $n$) | climb(M, $n+1$)}) \to_c t \ao \texttt{climb(M, $n+1$)} \]
does not contain a single system transition or $\opsem$ step. Naively, we could have defined the strategy as simply
\begin{lstlisting}
sd run(M) := matchrew S s.t. M' := append(M, S) by S using (
	match S s.t. accept(M') | all ; run(M')
) .
\end{lstlisting}
However, while representing the language $L_* = \{t\}^*$ for some term $t$, the infinite repetition of $t$ will be inevitably allowed because of the execution that always takes the second branch.

	The case of $\omega$-languages is more complicated, but the proof is similar. We assume that the language $L_\omega$ is represented by a nondeterministic Turing machine with Büchi conditions and type 2 semantics~\cite{omegaTuring}. They are defined as tuples $M = (Q, \Sigma, \Gamma, \delta, F, q_0)$ where $Q$ is a finite set of states, $\Sigma$ is a finite input alphabet (in our case, a subset of $T_\Sigma$), $\Gamma$ is a finite tape alphabet with $\Sigma \subseteq \Gamma$, $F$ is a set of states to define the Büchi condition, $q_0 \in Q$ is the initial state, and $\delta : Q \times \Gamma \to \mathcal P(Q \times \Gamma \times \{L, R, S\})$ is the nondeterministic transition function. A \emph{run} of $M$ for a word $w$ is an infinite sequence of configurations $r = (q_i, w_i, j_i)^\infty_{i=1}$ with $r_1 = (q_0, w, 0)$ and $r_{k+1} = (q_{k+1}, w_k[j_k/s], j_k + m)$ if $(q_{k+1}, s, m) \in \delta(q_k, (w_k)_{j_k})$ where $m$ is $-1$ for $L$, $1$ for $R$, and $0$ for $S$. A run is \emph{complete} if every position of the tape is ever visited. A word is \emph{accepted} if there is a run such that $q_i \in F$ infinitely often.

\begin{lstlisting}
sd climb(M, N) := run(M, N) | climb(M, s(N)) .
sd run(M, 0) := match S s.t. final(M) ; climb(M, 1) .
sd run(M, s(N)) := match S s.t. needsInput(M) ?
	  all ; matchrew S' by S' using run(put(M, S'), s(N))
	: matchrew S s.t. M', MS := step(M) by S using run(M', N) .
\end{lstlisting}
The \texttt{climb} definition is identical to the finite case, but now it fixes the next configuration where a final state of the Turing machine must be found, and it is called repeatedly to ensure that those are visited infinitely often. Since executing the Turing machine after writing an infinite word into the tape is not possible, we advance it while running the strategy and fill the tape lazily when required. When the machine moves right to a blank position, this is revealed by the \texttt{needsInput} predicate, a rewrite step is executed, and the new term is put in place of the blank before it can be read. Each step of the Turing machine consumes the counter and when it bumps into zero, the current state of the Turing machine is checked to be final. If it is not, the execution is discarded. Otherwise, a new call to \texttt{climb} ensures that a final state will be visited again.

	Let $\alpha$ be \texttt{climb(M0, 0)}. First, $E(\alpha) \subseteq L(M)$. If $w \in E(\alpha)$, by definition of $E(\alpha)$ and $\opsem$, there must be some $(q_n)_{k=0}^\infty \in \xs^\omega$ and $(n_k)_{k=0}^\infty \in \N^\omega$ such that $q_n \to_{s,c} q_{n+1}$, $q_{n_k} \opsem q_{n_{k+1}}$ and $\cterm(q_{n_k}) = w_k$. The only rule application in the strategies involved is the \skywd{all} in the positive branch of the conditional of the second \texttt{run} definition. Hence, this branch must have been executed infinitely many times, and so the machine must have moved its head infinitely many times to positions of the tape that need input. The machine is moved only in the negative branch of the same definition by a strategy-call $\to_c$ transition, so let $(m_k)_{k=0}^\infty$ be the indices of the states followed by these transitions. Taking the argument \texttt{M} of these calls, a run $(c_k)_{k=0}^\infty$ of the Turing machine can be constructed. In effect, $c_k \vdash c_{k+1}$ by the meaning of \texttt{move}, the run is complete since it visits infinitely many positions of the tape, and so the entire tape since it moves only one cell at a time, and the contents of the tape is the word $w$ since this is what \texttt{put} inserts each time a rule is executed. Moreover, the Büchi condition is satisfied because of the \texttt{climb} strategy: at any configuration $c_k$, the number of steps until a new final state is reached can be read from the second argument of the \texttt{run} call. In conclusion, $(c_k)_{k=0}^\infty$ is an accepting run of the machine, and so $(w_k)_{k=0}^\infty \in L(M)$.

	To prove the converse $L(M) \subseteq E(\alpha)$, let $(c_k)_{k=0}^\infty$ be a complete and accepting run of the Turing machine for some word $w \in L_\omega$. Since it is accepting, it must visit infinitely many final states, and there exists $(n_k)_{k=0}^\infty$ with $n_k \geq 1$ such that $c_{n_k}$ is in a final state. Moreover, since the run is complete, all the positions of the tape must be visited, so there is a $(m_k)_{k=0}^\infty$ such that the machine visits the position $k$ for the first time in $c_{m_k}$. With these ingredients, we can construct a nonterminating derivation of the operational semantics: starting at $q_0 = t_0 \ao \alpha$, \texttt{climb} calls \texttt{run} with $n_0$, and then the execution of \texttt{run} is deterministic until \texttt{N} reaches zero except for \skywd{all} and the selection of the move of the nondeterministic machine. Each time the second branch of the conditional has to be executed, we choose the next machine configuration $c_k$ in the run in the \skywd{matchrew}. Similarly, when the first branch is executed, the result of \skywd{all} is chosen to match the value of the current cell in $c_k$ and this is always possible since $(w_k)_{k=0}^\infty$ is a valid rewriting path of the uncontrolled system. When the counter descends to zero, the test in the \texttt{run} definition is satisfied, since the configuration is some final $c_{n_k}$, and the new \texttt{run} argument generated by \texttt{climb} is chosen to be $n_k - n_{k+1}$, and this procedure is repeated forever. The resulting $\to_{s,c}$ derivation contains infinitely many $\to_s$ transitions as a consequence of the completeness of the machine run, and so a $\opsem$ derivation can be extracted whose projection is the expected word $w$ since the \skywd{all} outputs have been chosen to match it. Therefore, $w \in E(\alpha)$.

\end{proof}

\begin{proposition}
	If the reachable states from $t \ao \alpha$ are finitely many, $E(\alpha, t)$ is a closed $\infty$-regular language.
\end{proposition}

\begin{proof}
	The Büchi automaton for $E(\alpha, t)$ is $A = (Q, \cterm(Q), \delta, \{\mathit{start}\}, Q)$ where $Q = \{\mathit{start}\} \cup \{ q \in \xs \mid t \in I \; \wedge \; t \ao \alpha \opsem^* q \} \cup \{ t' \ao \varepsilon : t \ao \alpha \to_c^* t' \ao \varepsilon \}$ and
\begin{align*}
	\delta(start, t) &= \{\; t \ao \alpha \;\} & \text{if } t \in I \\
	\delta(start, t) &= \emptyset & \text{if } t \not\in I \\
	\delta(q, t) &= \{\; q' : q \opsem q' \;\wedge\; \cterm(q') = t \;\} &\text{for } q \in Q \,\backslash\, \{\mathit{start}\} \\
		&\hphantom{= } \cup \{\; t \ao \varepsilon : \textrm{if } q \to_c^* t \ao \varepsilon \;\}
\end{align*}
The identity $L(A) = E(\alpha, I)$ follows from the fact that runs $\pi$ in $A$ yield executions $\cterm(\pi)$ in $E(\alpha, t)$ and vice versa. Proving this is straightforward, in the account of the definitions of $A$ and $E(\alpha, t)$. The proof for finite words is identical.

\end{proof}

\begin{lemma} \label{opal:existmatchrew}
	Given two terms $t,t' \in T_\Sigma$ such that $t \to^1_R t'$, there is a strategy expression $\alpha_{t,t'}$ of the form $\skywd{matchrew} \; P \; \skywd{by} \; x \; \skywd{using} \; rl[\rho]\{\bar \beta\}$ such that $t \ao \alpha_{t,t'} \to_{s,c}^* u \ao \varepsilon$ iff $t' = u$ and there are finitely many reachable states from $\alpha_{t,t'}$.
\end{lemma}

\begin{proof}

	Notice that the much simpler strategy \texttt{\skywd{all} ; \skywd{match} $t'$} also satisfies the first requirement, but not necessarily the second since the rewriting condition may have infinitely many solutions.
If $t \to^1_R t'$, there must exist a (perhaps conditional) rule $\mathit{rl} : l \to r \; \mathrm{if}\; C$, a substitution $\sigma$, and a position $p$ in $t$ such that $t|_p = \sigma(l)$, $t' = t[p/\sigma(r)]$ and $\sigma(C)$ holds. Proceeding by induction on the number of rewriting conditions required to prove a step, we first suppose that $C$ does not contain rewriting condition fragments. If $x_1, \ldots, x_n$ are the variables that occur in $l$ and $C$, and $x$ is a fresh variable, the desired $\alpha$ is then
\[ \skywd{matchrew} \; t[p\,/\,x] \; \skywd{by} \; x \; \skywd{using} \; \mathit{rl}\,[x_1 \leftarrow \sigma(x_1), \ldots, x_n \leftarrow \sigma(x_n)] \]
This strategy forces the $\mathit{rl}$ rule application to the specific position $p$, with the specific substitution $\sigma$. The only possible execution is
\begin{align*}
	t \ao \alpha_{t,t'} &\to_c \subterm(x : t|_p \ao \mathit{rl}\,[x_k \leftarrow \sigma(x_k)]_{k=1}^n; t[p\,/\,x]) \ao \varepsilon \\
	&\to_s \subterm(x : \sigma(r) \ao \varepsilon; t[p\,/\,x]) \ao \varepsilon \to_c t' \ao \varepsilon
\end{align*}

	If $C$ contains rewriting conditions, we have to indicate strategies for these. Since $t \to^1_R t'$, for each rewriting condition $l' \ttrew r'$, a sequence $t_1 t_2 \cdots t_n$ must exist with $\sigma(l') = t_1$, $\sigma(r') = t_n$ and $t_k \to^1_R t_{k+1}$. Some of these steps may apply rules with rewriting conditions, but we are one level less, so the existence of $\alpha_{t_k, t_{k+1}}$ can be assumed. Joining all the transitions with the concatenation operator of strategies, a strategy is built to solve one of the rewriting conditions. The same can be done for the other rewriting fragments, so the lemma holds.
\end{proof}

\begin{proposition}
	If $L$ is a closed $\infty$-regular language, there is a strategy expression $\beta$ such that $E(\beta) = L$ and the reachable states from $t \ao \beta$ are finitely many for all $t \in T_\Sigma$.
\end{proposition}

\begin{proof}
	The proofs for the finite and the infinite cases are similar, so only the infinite case is considered. Approximately, the strategy expression $\beta$ will be the translation of the $\omega$-regular expression for the language $L$. Since $L$ is $\omega$-regular, there must be a Büchi automaton $A = (Q, S, \delta, Q_0, F)$ for $L$. However, the symbols of the alphabet are states and our language is based on rules, so we have to translate it.
The translation $B = (S \times Q, RA, \Delta, B_0, S \times F)$ is defined using the strategies of~\cref{opal:existmatchrew}, $RA = \{ \alpha_{t,t'} \mid t, t' \in S \}$ with $\Delta((t, q), \alpha) = \{\; (t', q') \mid t \ao \alpha \to_{s,c}^* t' \ao \varepsilon, \; q' \in \delta(q, t') \;\}$ and $B_0 = \{\; (w_0, q) \mid w \in L, \; q \in \delta(q_0, t), \; q_0 \in Q_0 \;\}$. It is easy to prove that $\alpha_{t, t'}$ satisfies the definition of $\Delta$ for any pair of terms and that $L = \{ v \in T_\Sigma^\omega \mid v_k \ao w_k \to_{s,c}^* v_{k+1} \ao \varepsilon \text{ for all } k \in \N, w \in L(B)\}$.

	Since $L(B)$ is $\omega$-regular, it can be expressed as an $\omega$-regular expression \cite{infregexp}, which always have the form $r_1 s_1^\omega + \ldots r_n s_n^\omega$ for $r_i, s_i$ regular expressions and $\varepsilon \not\in L(s_i)$. The conversion from regular expressions to strategy expressions is almost an identity. $\emptyset$ is translated as $\fail$, $\varepsilon$ as $\idle$, alternation, concatenation, and iterations are the same in both languages. For each $s_i$, to represent $s_i^\omega$, we define a named strategy with label $f_i$ without argument and defined as $T(s_i) \seq f_i$ if $T$ is the translation function.

	Inductively, we will prove that any successful execution $t \ao T(r)$ for any regular or $\omega$-regular expression $r$ sequentially executes all strategies in a word $w \in L(r)$, and that all words in $L(r)$ can be successfully executed for some initial term. This implies, from what we have proved above, that the traces for $T(r)$ are exactly $L$ as we want to prove. Splitting the execution in $RA$ \emph{atoms} is always possible, since they are the only rule applications in the sequence enclosed by \skywd{matchrew} opening and closing transitions, with possibly some control steps between these atoms. The proof is by induction on the structure of regular and $\omega$-regular expressions. However, we should take care that the semantics of the iteration is different from that of the Kleene star, since the iteration body could be repeated indefinitely. Since $E(\alpha, t)$ is closed, this infinite execution will be already included, so it makes no difference.

	Finally, and since the strategy satisfies the conditions of the second statement of~\cref{prop:finiterec}, the reachable states are finite.

\end{proof}

\begin{proposition}
	The reachable states from $t \ao \alpha$ are finitely many if any of the following conditions holds:
	\begin{enumerate}
		\item $\alpha$ does not contain iterations or recursive calls.
		\item The reachable terms from $t \ao \alpha$ are finitely many and all recursive calls in $\alpha$ and the reachable strategy definitions are tail.
	\end{enumerate}
\end{proposition}

\begin{proof}
	The first statement can be proved by induction on the execution states ranked by the lexicographic combination of the number of strategy constructors in their stacks, the number of execution state constructors, and the number of condition fragments in $\rewcond$ states. Looking at the rules, each possible execution state has a finite number of successors by the $\to_{s,c}$ relation, and the induction hypothesis can be applied for all but iterations and calls. In case no recursive strategies are called, reachable states can be proved finite by induction on the finite and acyclic static call graph.

	For the second statement, we know that the number of terms that can appear either as subject or as strategy call arguments in execution states is finite, so iteration and tail-recursive calls can be handled. The body $\alpha$ of an iteration $\alpha \texttt*$ cannot contain recursive strategy calls, because it would not be tail calls. Inductively on the number of nested iterations, there are finitely many reachable states from $t \ao \alpha \texttt* \, s$ in addition to those from the reachable $t' \ao s$ states at the end of the iteration. Assuming there are no iterations in $\alpha$, the only successors of that state are $t \ao \alpha \, \alpha \texttt* \, s$ and $t \ao s$. The reachable states from $t \ao \alpha \, \alpha \texttt* \, s$ are those reachable from $t \ao \alpha \, \vctx(s)$ with $\vctx(s)$ replaced by $\alpha \texttt* \, s$, and those reachable from $t' \ao \alpha \texttt* \, s$ for each solution yield by $t \ao \alpha \, \vctx(s)$. The first are finitely many by the first statement, and the second case is the same we are proving now regardless of the particular $t$ or $t'$, which are finitely many. Hence, the reachable states before $s$ are finitely many, and the same can be proven if $\alpha$ contains iterations by continuing the induction.

	Consider now an execution state $t \ao \mathit{sl}\,(t_1, \ldots, t_n) \, \sigma \, s$ where $\mathit{sl}$ is a recursive strategy. Its successors are $t \ao \delta \, \sigma' \, s$ for some definition $\delta$ and substitution $\sigma'$. If the expression $\delta$ does not contain recursive strategies, finitely many states are reachable from $t \ao \delta \, \sigma'$. Otherwise, all recursive calls are tail and this yields finitely many states plus some $t_k \ao \mathit{sl}_k(t_{k,1}, \ldots, t_{k,n_k}) \, \sigma' \, s$ and their successors. More precisely, it should be proved that our syntactical definition of tail call ensures this, but it is a straightforward inductive check. Since the possible $t_k$, $t_{k, l}$, and $\mathit{sl}_k$ are finitely many, the successors of initial state before reducing $s$ are finitely many, and combining all the results the whole reachable states are a finite set.
\end{proof}
 
\end{document}